\documentclass[a4paper,usenatbib]{mnras}

\usepackage[T1]{fontenc}
\usepackage{ae,aecompl}

\usepackage{graphicx}	
\usepackage{amsmath}	
\usepackage{amssymb}	
\usepackage{textcomp}
\usepackage{upgreek}
\usepackage{hyperref}
\usepackage{breakurl}
\usepackage{soul}

\newcommand{\urlwofont}[1]{\urlstyle{same}\url{#1}}
\newcommand{\kms}{km~s$^{-1}$}

\title[SN~2016gsd]{SN~2016gsd:  An unusually luminous and linear type II supernova with high velocities}

\author[T. M. Reynolds et al.]{T. M. Reynolds,$^{1}$\thanks{E$-$mail: thmire@utu.fi}
M. Fraser,$^{2}$
S. Mattila,$^{1}$
M. Ergon,$^{3}$
L. Dessart,$^{4}$ \newauthor
P. Lundqvist,$^{3}$ 
Subo Dong,$^{5}$
N. Elias-Rosa,$^{6,7}$ 
L. Galbany,$^{8}$ 
C. P. Guti\'errez,$^{9}$ \newauthor
T. Kangas,$^{10}$ 
E. Kankare,$^{1}$ 
R. Kotak,$^{1}$ 
H. Kuncarayakti,$^{1,11}$
A. Pastorello,$^{12}$ \newauthor
O. Rodriguez, $^{13,14}$
S. J. Smartt,$^{15}$
M. Stritzinger,$^{16}$
L. Tomasella,$^{12}$
Ping Chen,$^{5}$ \newauthor
J. Harmanen$^{1}$
G. Hosseinzadeh,$^{17}$ 
D. Andrew Howell,$^{18,19}$
C. Inserra,$^{20}$ 
M. Nicholl,$^{21}$ \newauthor
M. Nielsen,$^{16}$ 
K. Smith,$^{15}$ 
A. Somero,$^{1}$ 
R. Tronsgaard,$^{22}$ 
D. R. Young$^{15}$
\\
$^{1}$Tuorla Observatory, Department of Physics and Astronomy, University of Turku, FI$-$20014, Finland\\
$^{2}$School of Physics, O'Brien Centre for Science North, University College Dublin, Belfield, Dublin 4, Ireland.\\
$^{3}$Oskar Klein Centre, Department of Astronomy, AlbaNova, Stockholm University, Stockholm, Sweden \\
$^{4}$Unidad Mixta Internacional Franco-Chilena de Astronom\'{i}a (CNRS, UMI 3386), Departamento de Astronom\'{i}a, Universidad de Chile, \\ Camino El Observatorio 1515, Las Condes, Santiago, Chile \\
$^{5}$Kavli Institute for Astronomy and Astrophysics, Peking University, Yi He Yuan Road 5, Hai Dian District, Beijing 100871, China \\
$^{6}$Institute of Space Sciences (ICE, CSIC), Campus UAB, Carrer de Can Magrans s/n, 08193 Barcelona, Spain \\
$^{7}$Institut dEstudis Espacials de Catalunya (IEEC), c/Gran Capit \'{a} 2$-$4, Edif.  Nexus 201, 08034 Barcelona, Spain \\
$^{8}$Departamento de F\'isica Te\'orica y del Cosmos, Universidad de Granada, E-18071 Granada, Spain. \\
$^{9}$Department of Physics and Astronomy, University of Southampton, Southampton, SO17 1BJ, UK \\
$^{10}$Space Telescope Science Institute, 3700 San Martin Drive, Baltimore, MD 21218, USA \\
$^{11}$Finnish Centre for Astronomy with ESO (FINCA), FI-20014 University of Turku, Finland. \\
$^{12}$INAF $-$ Osservatorio Astronomico di Padova, Vicolo dell'Osservatorio 5, 35122 Padova, Italy \\
$^{13}$Departamento de Ciencias Fisicas, Universidad Andres Bello, Avda. Republica 252, Santiago, Chile \\
$^{14}$Millennium Institute of Astrophysics (MAS), Nuncio Monse\~nor S\'otero Sanz 100, Providencia, Santiago, Chile \\
$^{15}$Astrophysics Research Centre, School of Mathematics and Physics, Queen's University Belfast, BT7 1NN, UK \\
$^{16}$ Department of Physics and Astronomy, Aarhus University, Ny Munkegade 120, DK$-$8000 Aarhus C, Denmark \\
$^{17}$Center for Astrophysics \textbar{} Harvard $\&$ Smithsonian, 60 Garden Street, Cambridge, MA 02138-1516, USA \\
$^{18}$Las Cumbres Observatory, 6740 Cortona Dr Ste 102, Goleta, CA 93117$-$5575, USA \\
$^{19}$Department of Physics, University of California, Santa Barbara, CA 93106$-$9530, USA \\
$^{20}$ School of Physics \& Astronomy, Cardiff University, Queens Buildings, The Parade, Cardiff, CF24 3AA, UK \\
$^{21}$Institute for Astronomy, University of Edinburgh, Royal Observatory, Blackford Hill, EH9 3HJ, UK \\
$^{22}$DTU Space, National Space Institute, Technical University of Denmark, Elektrovej 328, DK-2800 Kgs. Lyngby, Denmark
}
\date{Accepted 2020 February 5. Received 2020 February 5; in original form 2019 September 12}
\pubyear{2019}

\begin{document}
\label{firstpage}
\pagerange{\pageref{firstpage}$-$\pageref{lastpage}}
\maketitle

\begin{abstract}
We present observations of the unusually luminous type II supernova (SN) 2016gsd. With a peak absolute magnitude of $V=-$19.95$\pm$0.08, this object is one of the brightest type II SNe, and lies in the gap of magnitudes between the majority of type II SNe and the superluminous SNe. Its light curve shows little evidence of the expected drop from the optically thick phase to the radioactively powered tail. The velocities derived from the absorption in H$\alpha$ are also unusually high with the blue edge tracing the fastest moving gas initially at 20000 \kms~and then declining approximately linearly to 15000 \kms~over $\sim$100 days. The dwarf host galaxy of the SN indicates a low metallicity progenitor which may also contribute to the weakness of the metal lines in its spectra. We examine SN~2016gsd with reference to similarly luminous, linear type II SNe such as SNe 1979C and 1998S, and discuss the interpretation of its observational characteristics. We compare the observations with a model produced by the {\sc jekyll} code and find that a massive star with a depleted and inflated hydrogen envelope struggles to reproduce the high luminosity and extreme linearity of SN~2016gsd. Instead, we suggest that the influence of interaction between the SN ejecta and circumstellar material can explain the majority of the observed properties of the SN. The high velocities and strong H$\alpha$ absorption present throughout the evolution of the SN may imply a CSM configured in an asymmetric geometry.

\end{abstract}




\section{Introduction}
Hydrogen$-$rich (H-rich) core$-$collapse supernovae (SNe) come from massive stars that have retained a significant H-rich envelope until the end of their lives, and are observationally classified on the basis of broad Balmer lines in their spectra \citep{Filippenko1997}. Such ``type II'' SNe have often been divided into two classes: type IIP SNe that show a plateau in their lightcurves, and type IIL SNe that follow a linear decline from peak \citep{Barbon1979}. There are also two further classes of SNe that exhibit H lines in their spectra, type IIn SNe that display narrow lines of H associated with interaction between the SN ejecta and the progenitor circumstellar medium (CSM) \citep{Schlegel1990}  and type IIb SNe, which show broad H lines only at early times and broad He lines later on (see \citet{Galyam2017} and references therein).

The distinction between type IIP and IIL SNe found support from studies based on small samples of observed events \citep{Patat1993,Patat1994,Arcavi2012,Faran2014b}. Recently though, the number of type II SNe observed has increased due to wide field surveys, and studies with sample size greater than a few dozen objects \citep[e.g.][]{Anderson2014a,Sanders2015,Valenti2016,Galbany2016,Rubin2016} do not find that the light curves of type II SNe alone are sufficient to make this distinction. These and other studies have confirmed that type II SNe with a less pronounced plateau are typically brighter than type IIP SNe and that their spectra show higher velocities and weaker H features at $\sim$50 days \citep[see e.g.][]{Faran2014b,Gutierrez2014,Anderson2014b,Valenti2015,Sanders2015}. In this work we will use the IIL/P nomenclature both to refer to SNe that have been published under this classification system and to refer to objects at either end of the range of linearity observed in type II light curves, while noting that there is no unambiguous division within this population.

The progenitor systems for type IIP SNe have been firmly established as red supergiant (RSG) stars with zero$-$age main$-$sequence (ZAMS) masses between 8 and $\sim$18 $M_{\odot}$ through direct identification in pre$-$SN images \citep[e.g.][]{VanDyk2003,Smartt2004,Smartt2015a}. It is expected from theory that type IIP events are caused by RSGs undergoing Fe core$-$collapse, with the characteristic plateau phase being associated with H recombination within the H-rich envelope once the photospheric temperature reaches $\sim$6000K. During the plateau phase, the recession velocity of the photosphere is similar to the expansion velocity of the ejecta, causing an approximately constant brightness.
Once the shock deposited energy within the envelope is depleted, there is a rapid drop in the SN light curve from the plateau phase onto the ``radioactive tail'' phase. Here the luminosity is powered by the radioactive decay of $^{56}$Co $\rightarrow$ $^{56}$Fe. This has been successfully modeled for the assumed progenitor systems by a number of authors \citep[see e.g.][]{Grassberg1971,Falk1977,Kasen2009, Dessart2011,Dess13}. However, the light curve can not alone be used to estimate the progenitor mass, as modelling shows that a number of different explosion parameters can reproduce similar light curves \citep{Dessart2019b,Goldberg2019}.

From this understanding of type IIP SNe, it has been inferred that type IIL SNe could arise from progenitor stars with a less massive H$-$rich envelope. In this case, with less H$-$rich material, the optical depth within the H$-$rich envelope is insufficient to form a plateau phase in the light curve, and the stored radiation is released more quickly \citep{Grassberg1971,Blinnikov1993}. This can be supported by the weaker H absorption observed in type IIL SNe, which is explained by a lower column density of absorbing material. To form such systems, the progenitor star must have had much of its envelope removed through stellar winds, or through interaction with a binary companion. Type IIL SNe are also relatively rare, making up $\sim$7.5\% of the volume limited core$-$collapse SN sample of \cite{Li2011}, and show higher average brightness and velocities (at $\sim$50 days) than type IIP SNe as mentioned above. This leads to the hypothesis that type IIL SNe could arise from relatively higher mass progenitor systems which would be less common, have higher explosion energy and have more extreme mass loss histories. 

There are very few direct observational constraints for the progenitors of type IIL events. The two type IIL SN for which pre$-$explosion imaging is available did not provide unambiguous detections of their progenitors, although they indicated upper limits of < 20 M$_{\odot}$ in the case of SN 2009hd \citep{Elias-Rosa2011} and $<25$ M$_{\odot}$ in the case of SN 2009kr \citep{Maund2015}. \cite{Smartt2015a} find a statistically significant lack of high mass (>18 M$_{\odot}$) RSG progenitors for type IIP events, which implies that SNe from higher mass stars are usually of a different type, if they explode at all. However, there is ongoing discussion about this result \citep[see e.g.][]{Martinez2019,Davies2018,Walmswell2012}.

Another method of constraining the progenitor systems of SNe is by analysing spectra taken at late times (>200 days), when the ejecta have expanded and become optically thin, allowing observation deep into the ejecta \citep[e.g.][]{Fransson1989,Jerkstrand2012}. In these nebular spectra we observe emission from products of late stage nucleosynthesis in the progenitor star, the yields of which are sensitive to the progenitor ZAMS mass \citep{Woosley2007}. \cite{Jerkstrand2014} found, from the strength of the [O {\sc i}] line in observations, no evidence that any type IIP progenitor was more massive than 20 M$_{\odot}$. Using the same method of analysis, on a sample of 12 IIL SNe, \cite{Valenti2016} also find no evidence for high mass progenitors, although this is under the assumption that fundamental parameters such as explosion energy and mass loss history are similar for IIP and IIL SNe. However, \citet{Anderson2018} suggested that the type IIP SN~2015bs had a progenitor with mass 17$-$25M$_{\odot}$ based on the strongest [O {\sc i}] emission feature yet seen in a type II SN.

A further way to constrain the progenitor systems of SNe is to consider their explosion sites. \cite{Kuncarayakti2013} attempted to compare the explosion sites of type IIP and type IIL SNe with integral field unit (IFU) spectroscopy and found that IIL SNe come from younger environments, although they only considered 19 sites, of which 3 were type IIL SN sites. More recently, \cite{Galbany2018} found a connection between IIL SNe and high local star formation rates in a much larger sample of 95 type II SNe. \cite{Kangas2017} found that more massive stars are connected with type IIb and IIL SNe than with type IIP SNe through correlation of stellar populations with H$\alpha$ emission.

Given the assumption of a enhanced mass loss history to explain the diminished H$-$rich envelope of type IILs, it is unsurprising that there are signs of interaction between the SN ejecta and CSM in many type IIL SNe. The prototypical type IIL SN 1979C \citep{Branch1981,DeVaucouleurs1981a} showed a number of signatures of high mass loss rate, including its radio light curves \citep{Lundqvist1988,Weiler1991}, X-ray emission \citep{Immler2005} and a near$-$IR (NIR) echo from a circumstellar dust shell \citep{Dwek1983}. It was also abnormally bright for a type IIL \citep[see][]{Richardson2014}, which could be explained by interaction. The most commonly considered sign of interaction in SN observations is the presence of persistent narrow lines in the Balmer series, as observed in type IIn SNe. These lines are often only present at early times before the densest CSM material is swept up by the SN ejecta. This can be observed in the case of SN~1998S, which was spectroscopically classified as a type IIn based on narrow lines in the early spectra \citep{Liu2001,Fassia2000,Fassia2001}. Its later spectra and light curves look very much like those of SN~1979C, whose earliest spectra show no clear evidence of narrow lines.  More recently, narrow lines at early times have been observed in multiple objects that also exhibit the characteristic brightness, light curve shape and spectral features of type IILs $-$ e.g. SN~2007pk \citep{Inserra2013},  SN~2008fq \citep{Taddia2013}, SN~2013by \citep{Valenti2015}. 
Other bright type IIL events such as SN 2013fc \citep{Kangas2016}, which peaked at B band magnitude $-$20.46 mag, show a blue continuum still at epochs of 3$-$4 weeks from the explosion, later than seen in a typical type II SN. This can also be attributed to the interaction of the SN ejecta with dense pre$-$existing CSM close to the progenitor \citep{Hillier2019}.

\begin{figure*}
\includegraphics[width=1\textwidth]{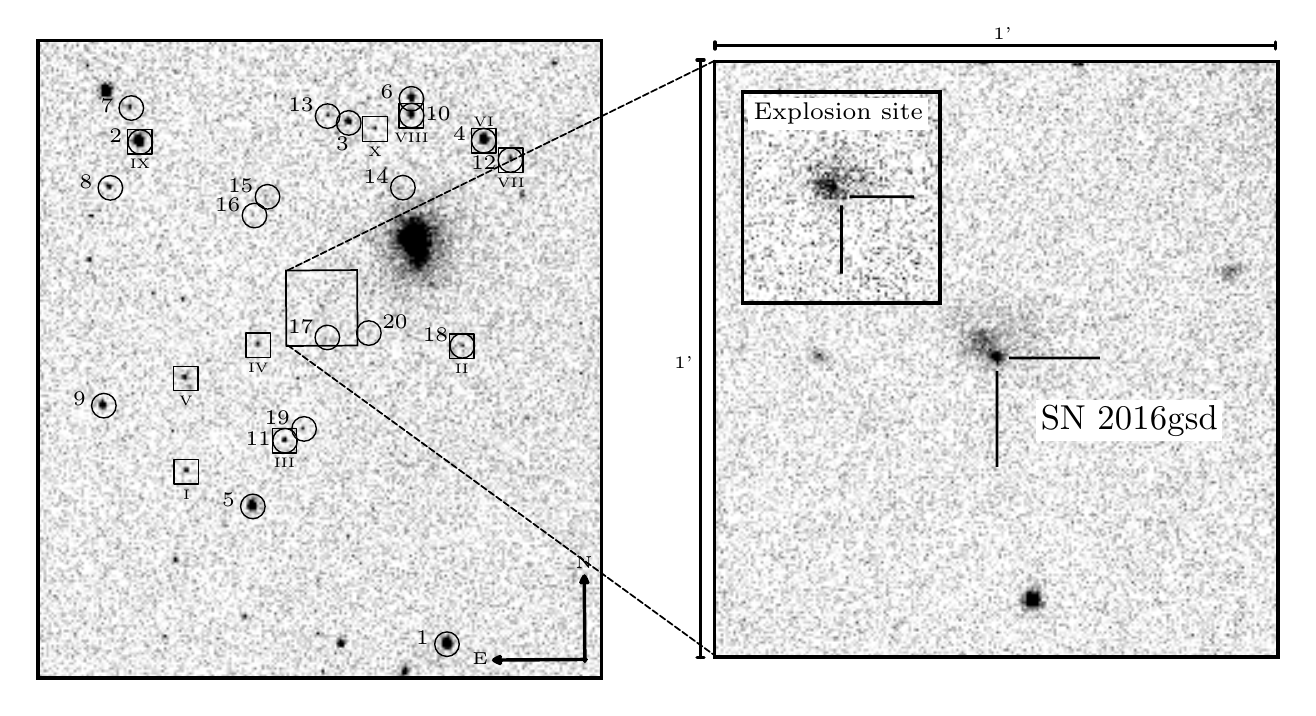}
\caption{Finding chart in $V$ band for SN~2016gsd. Optical calibration stars listed in Tab. \ref{tab:calibration} are circled, NIR calibration stars are marked by squares. The right hand side panel shows the SN location. The inset image shows where the SN exploded in our template image, taken after the SN has faded.}
\label{fig:finder}
\end{figure*}

In this work we consider SN~2016gsd, which is also exceptionally bright for a Type IIL SN and follows a long linear decline. In contrast to these other bright type IIL SNe, it shows unusually high velocities in its strong H P-Cygni absorption features and occurred in a low-luminosity host galaxy. The structure of the paper is as follows: in Section \ref{sect:ObsDataRed} we summarise the observational data and the data reduction process, in Sections \ref{sect:Phot} and \ref{sect:spec} we present and discuss the spectroscopy and photometry of SN~2016gsd respectively with associated analysis and in Section \ref{sect:Host} we discuss the host galaxy of the SN. In Section \ref{sect:Discussion}   we explore first if our observations can be reproduced by a model assuming a progenitor with a depleted and inflated H-rich envelope and afterwards whether a scenario with significant interaction between the SN ejecta and dense CSM is more appropriate. Finally in Section \ref{sect:Conclusions} we present our conclusions.

\section{Observations and data reduction} 
\label{sect:ObsDataRed}
\subsection{Discovery, classification and adopted parameters}
\label{sect:Discovery} 

\begin{figure*}
\includegraphics[width=1\textwidth]{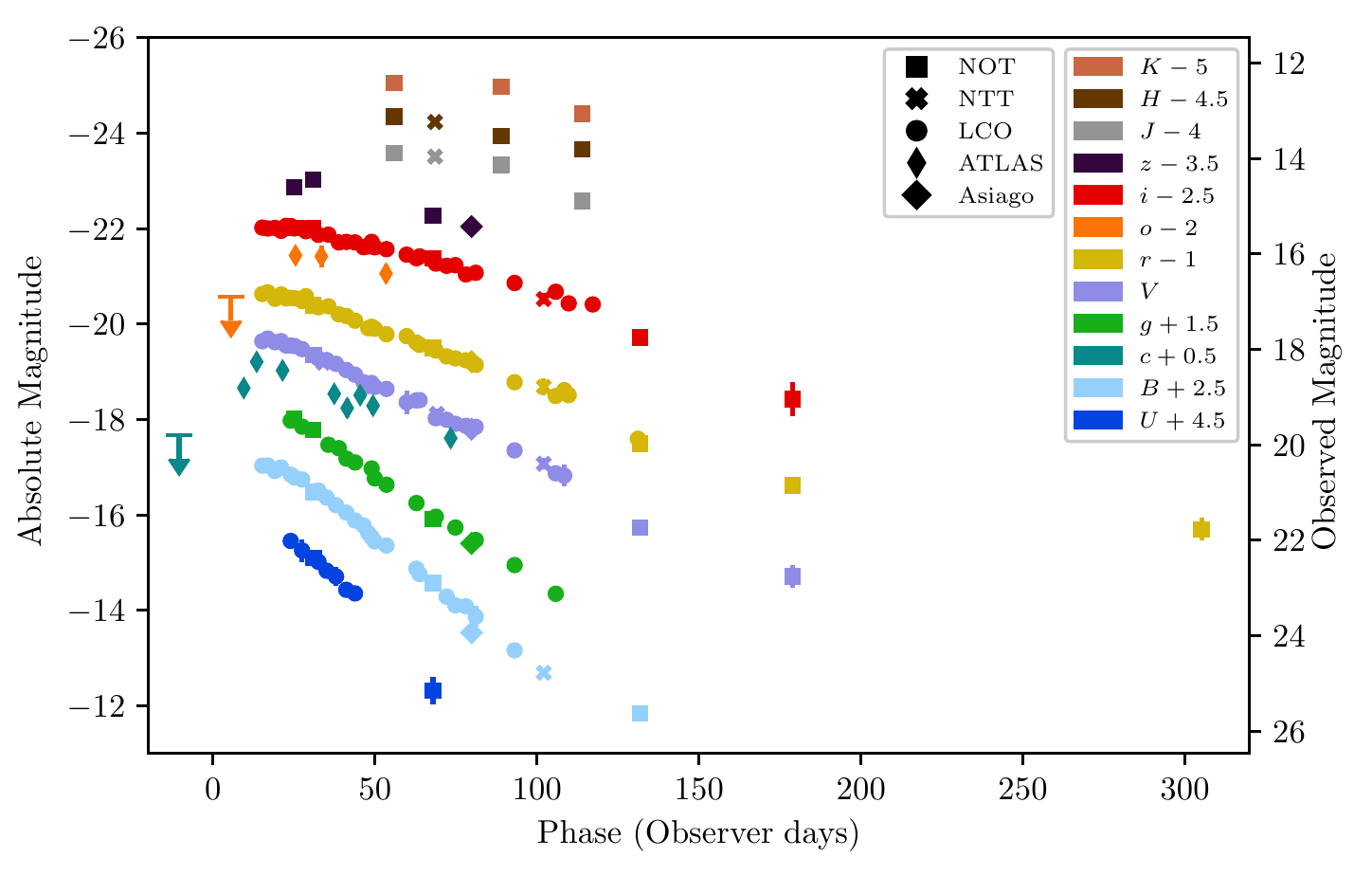}
\caption{Photometric light curves in \textit{UBgVrizJHK} and ATLAS \textit{c} and \textit{o} bands, in the observer frame, relative to our adopted explosion date (see Sect. \ref{subsec:Explosion epoch}). The data have been vertically shifted for visual clarity. The values given on the left axis are corrected for distance and not for Milky Way extinction. In filters for which we have only a handful of observations, namely Johnson$-$Cousins \textit{R}, \textit{I} and SDSS \textit{u} band, a conversion was made to a filter with more observations, also for visual clarity.}
\label{fig:lc}
\end{figure*}

SN~2016gsd was reported by K. Itagaki at an unfiltered magnitude of 17.7 mag on 2016 Sept. 29.7 \citep{Itagaki2016}, and subsequently classified by the NOT Unbiased Transient Survey (NUTS\footnote{http://csp2.lco.cl/not/}) as a luminous type II SN on Oct. 10.1 \citep{Toma16}.
SN~2016gsd is located at $\alpha = 02^{h}40^{m}34^{s}.44,~\delta =19^{\circ}16'59".90$, and offset by 1.03''W and 1.2''S relative to the nucleus of its faint host galaxy. This galaxy was previously catalogued by the Pan-STARRS 3Pi survey \citep{Chambers2016,Flewelling2016} as PSO J040.1438+19.28236 with a Kron r-band magnitude of $20.70 \pm0.05$ mag. The field of SN~2016gsd is shown in Fig. \ref{fig:finder}.
In the following, we adopt a redshift of $z$=0.067$\pm$0.001 towards SN~2016gsd (see Sect. \ref{sect:Host}). The corresponding luminosity distance and distance modulus are $D_{L} = 311.6$ Mpc and $\mu = 37.47$ mag respectively, assuming a flat cosmology with $H_{0} = 67.7~\mathrm{km~s}^{-1}~\mathrm{Mpc}^{-1}$ and $\Omega_{m} = 0.309$ \citep{Planck2016}. At this distance, the projected offset of SN~2016gsd from the host nucleus is 2.4 kpc.
We take the foreground extinction towards SN~2016gsd to be $A_V=0.248\pm0.004$ mag \cite[from][via the NASA Extragalactic Database, NED]{Schl11}. As discussed in Sect. \ref{sect:Phot}, we find the internal extinction in the host of SN~2016gsd to be negligible.
In the following, we adopt  MJD~$= 57648_{-4}^{+8}$ as the explosion date and the reference date for our imaging and spectroscopic epochs (see Sect. \ref{subsec:Explosion epoch}).

\subsection{Photometry}
\label{sect:Photometry}

Optical imaging of SN~2016gsd was obtained as part of the Las Cumbres Observatory (LCO, see \citet{Brown2013}) SN key project with their network of 1.0m telescopes, equipped with Sinistro cameras; with the 2.56m Nordic Optical Telescope (NOT\footnote{http://www.not.iac.es}), using the ALFOSC camera; with the 3.6m New Technology Telescope (NTT) equipped with the EFOSC2 camera; and with the 1.82 m Copernico Telescope using AFOSC. Processing for bias and flat field corrections was performed automatically through the {\sc banzai} pipeline for the data from LCO and with the {\sc alfoscgui}\footnote{{\sc foscgui} is a graphical user interface aimed at extracting SN spectroscopy and photometry obtained with FOSC$-$like instruments. It was developed by E. Cappellaro. A package description can be found at http://sngroup.oapd.inaf.it/foscgui.html.} software in the case of the ALFOSC data.  The EFOSC2 images  were  reduced  using  the  PESSTO  pipeline,  as  described  in \cite{Smartt2015b}, and the AFOSC images were reduced using standard tasks in {\sc iraf}. The SN magnitudes were obtained using the SNOoPy\footnote{SNOoPy is a python package for SN photometry using PSF fitting and/or template subtraction developed by E. Cappellaro. A package description can be found at http://sngroup.oapd.inaf.it/snoopy.html.} pipeline. 

\begin{figure*}
\includegraphics[width=1\textwidth]{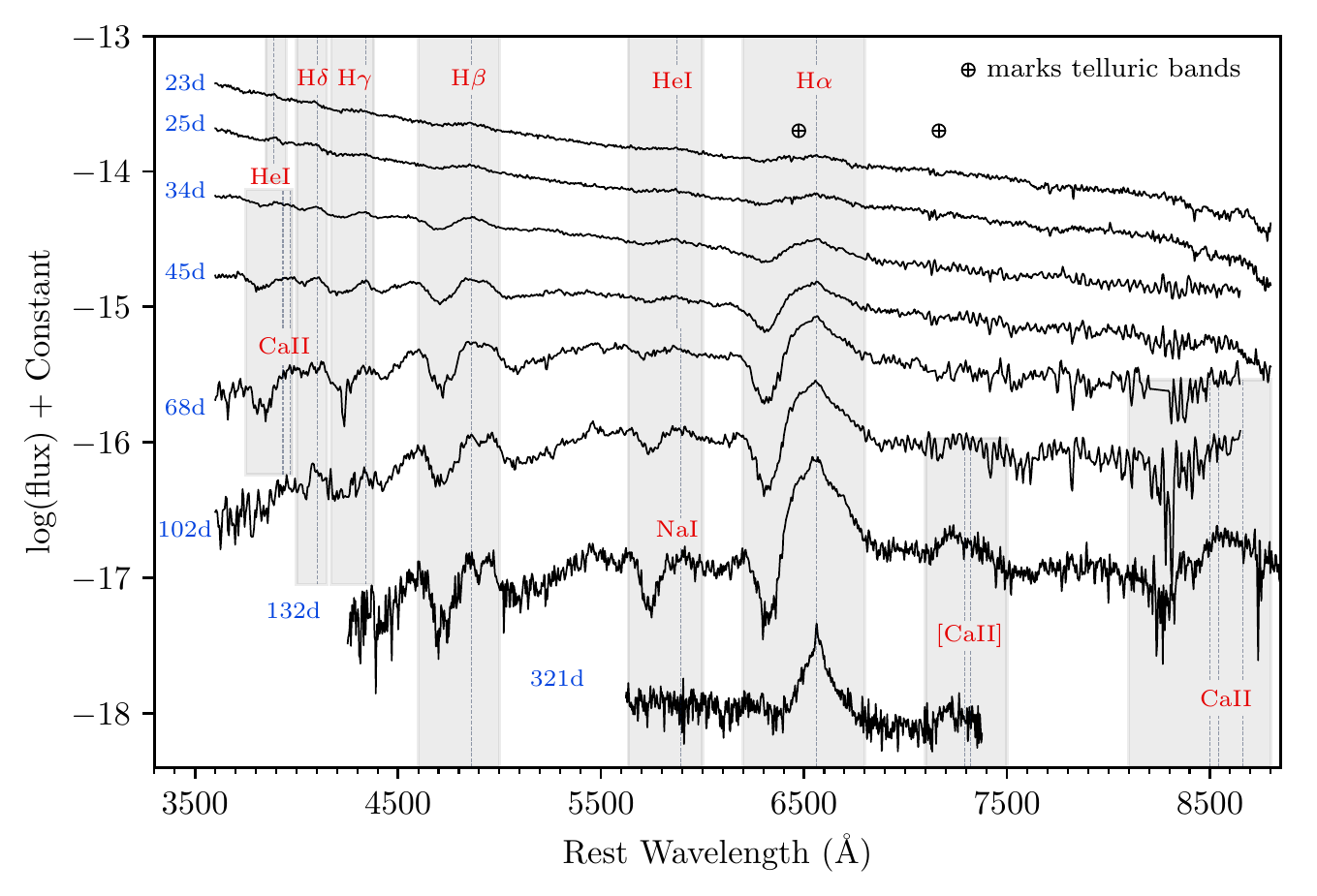}
\caption{Spectral sequence for SN~2016gsd. The pairs of spectra taken at +44 and +46 days and +67 and +69 days have been combined to improve the signal to noise ratio and are labeled with the mean date; spectra have been shifted vertically for clarity. The strongest features as discussed in the text are identified. Shaded regions show the approximate areas where the marked spectral features are present.}
\label{fig:spec_seq}
\end{figure*}

To obtain the instrumental magnitude of the SN via PSF fitting photometry, bright isolated field stars were selected to create a PSF model, simultaneously fitting the source and background for these field stars and iterating for the best fit. The SN magnitude was then obtained by fitting the PSF model to the SN while subtracting the nearby background estimated through a 2nd order polynomial fit. An uncertainty estimate is obtained through an artificial star experiment where a fake source is injected and recovered at multiple positions close to the SN. The dispersion of recovered artificial star magnitudes is combined in quadrature with the PSF fit error. The proximity of the SN to the centre of the host galaxy necessitated template subtraction at later times when the flux of the galaxy became a significant proportion of the SN flux, to remove the contaminating flux. Templates were taken at late times (+480d after our adopted explosion date), after the SN had faded, with better depth and seeing than the SN images. They were aligned to match the SN data, and subtracted using {\sc hotpants}\footnote{ {\sc hotpants} is available at https://github.com/acbecker/hotpants}, an implementation of the image subtraction algorithm by \cite{Alard1998}. Photometry was then performed in the subtracted image as before, with the appropriate PSF model. To determine when template subtraction was necessary, we checked for each filter if the template subtracted measurements were not consistent with the un-subtracted measurements. We report the template subtracted measurement for all epochs after the first epoch where the measurements were inconsistent, and measurements from the un-subtracted images for the earlier epochs. Template subtraction was not possible for our final photometric point in r band at +305d, as we were enable to get a subtraction of sufficient quality to measure the magnitude without host contamination. The listed magnitude makes use of the aforementioned background fit to attempt to subtract the galaxy flux.

To calibrate the zeropoints of the images taken in the \textit{griz} Sloan filters, we used aperture photometry to evaluate the instrumental magnitudes of multiple bright isolated field stars and compared these to values taken from the Pan-STARRS1 catalogue \citep{Chambers2016}. For images taken in the Sloan \textit{u} band, we performed the same procedure with values taken from the South Galactic Cap \textit{u}$-$band Sky Survey (SCUSS) \citep{SCUSS2016}. For images taken in the Johnson$-$Cousins filters, we converted the magnitudes of the same field stars from the PS$-$1 filter system to the Johnson$-$Cousins system using the equations published in \cite{Tonry2012}. For \textit{U} band, we used an equation published in \cite{Jordi2006}. We chose to do this as there were no available standard observations taken on photometric nights. We then compared the instrumental magnitudes to these converted values to evaluate the zeropoints in these images, sigma clipping to remove outlier values. The chosen field stars are shown in Fig. \ref{fig:finder}.  Magnitudes for field stars are given in Tab. \ref{tab:calibration}. The optical magnitudes of SN~2016gsd are listed in Tab. \ref{tab:UBVRI_phot} and Tab. \ref{tab:ugriz_phot}. The uncertainties listed are the sum in quadrature of the uncertainty given by the SNOoPy pipeline and the standard error of mean of the individual zeropoint measurements in the field of SN~2016gsd. All the photometric data are presented in Fig. \ref{fig:lc}. In bands where there are only 1$-$2 observations, magnitudes are converted to filters with more epochs of data for visual clarity. 

Imaging of the site of SN~2016gsd was also performed as a part of the ATLAS survey \citep[see][]{Stalder2017,Tonry2018}. Forced photometry at the position of SN~2016gsd was performed on template$-$subtracted images using the ATLAS pipeline. As multiple images were taken in each filter on each night, the average of these is reported in Tab. \ref{tab:ATLAS_phot}. The {\it c} and {\it o}$-$filters used by ATLAS correspond to {\it g}+{\it r} and {\it r}+{\it i} respectively, and we have not attempted to calibrate these to a standard system. 3$\sigma$ upper limits are reported in Tab. \ref{tab:ATLAS_phot} for epochs when the SN was not detected.

Near$-$infrared (NIR) imaging of SN$~$2016gsd was performed with the NOT using the NOTCam instrument and with the NTT using the SOFI instrument. The NOTCam data were reduced using a slightly modified version of the NOTCam Quicklook v2.5 reduction package\footnote{http://www.not.iac.es/instruments/notcam/guide/observe.html}. The SOFI data were reduced with the SOFI pipeline in the PESSTO package. The reduction process included differential flat$-$field correction, distortion correction, bad pixel masking, sky subtraction and stacking of the dithered images. Magnitudes for the SN were obtained with the same process as described above for the optical data. Calibration of the zeropoints for the data was performed through comparison of the magnitudes of isolated field stars with those available in  the Two Micron All Sky Survey (2MASS) catalogue. The SN magnitudes obtained are reported in Tab. \ref{tab:JHK_phot}.
\subsection{Spectroscopy}
\label{sect:Spectroscopy}

A log of spectroscopic observations of SN$~$2016gsd is reported in Tab. \ref{tab:Spectral log}, and the full spectral sequence is shown in Fig. \ref{fig:spec_seq}.

ALFOSC spectra were reduced using the {\sc alfoscgui} package. This uses standard {\sc iraf} tasks to perform overscan, bias and flat$-$field corrections as well as removal of cosmic ray artifacts using  {\sc lacosmic}  \citep{VanD01}. Extraction of the one$-$dimensional spectra was performed with the {\sc apall} task and wavelength calibration was done by comparison with arc lamps and corrected if necessary by measurement of skylines. The spectra were flux calibrated against photometric standard stars observed on the same night. The EFOSC2 spectra were reduced in a similar fashion to the ALFOSC data, using the PESSTO pipeline \citep{Smartt2015b}. The only difference was that for EFOSC2 spectra, a master sensitivity curve which was created from multiple standard stars observed on different nights, was used to flux calibrate the spectra. 
The OSIRIS spectra were similarly reduced with standard {\sc iraf} tasks. Spectra taken with the same instrument and on the same night have been co$-$added to improve their signal$-$to$-$noise.

Strong telluric absorption bands were removed using either a model (in the case of EFOSC2 data) or observed  (for ALFOSC and OSIRIS) telluric absorption spectrum, obtained from observations of smooth continuum spectrophotometric standards. At the redshift of SN~2016gsd the telluric B band (between 6860$-$6890\r{A}) unfortunately falls in the same region as the P$-$Cygni feature seen in H$\alpha$. We hence took care in scaling the telluric absorption spectrum to match the stronger A band, which lies in a clean region of continuum. The absolute calibration of the spectra was done by performing synthetic photometry on the spectra with \textit{BVgr} filters that matched those used to obtain the data. The fluxes obtained were multiplicatively scaled to match the fluxes obtained through photometric observations, and the average value of these was applied to the spectrum.

We also obtained a single spectroscopic observation with the FIRE spectrometer \citep{Simcoe2013} mounted on the 6.5m Magellan Baade telescope. The target was observed in echelle mode and the data was reduced with a reduction pipeline developed for the FIRE instrument\footnote{http://web.mit.edu/\~{}rsimcoe/www/FIRE/ob\_data.htm}.

A deep spectrum of the host galaxy was obtained on 2019-09-01 using the FORS2 instrument mounted on UT1 of the ESO VLT with a 1.3" slit and in parallactic angle. The spectrum was reduced using the FORS2 recipe with the ESO esoreflex pipeline \citep{Freudling2013}.

The set of spectroscopic data, along with the photometric tables, will be released through the Weizmann Interactive Supernova data REPository \citep[WISEREP\footnote{https://wiserep.weizmann.ac.il};][]{Yaron2012}. 


\section{Photometric analysis}
\label{sect:Phot}

\begin{figure*}
\includegraphics[width=1\textwidth]{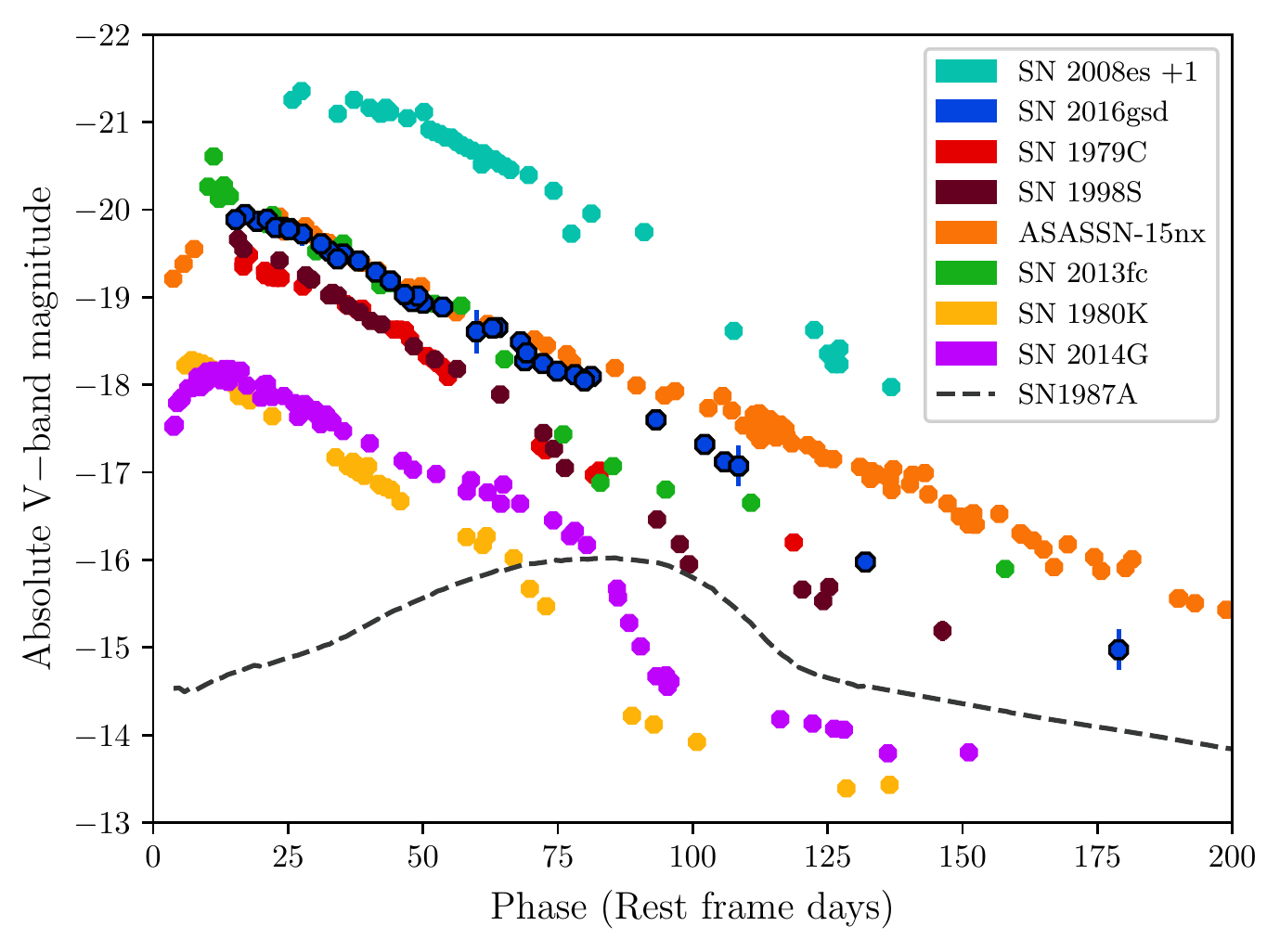}
\caption{The absolute V$-$band light curve of SN~2016gsd compared to other type II SNe. The light curve of SN 2008es has been shifted down by one magnitude. The adopted explosion date in MJD, distance in Mpc, E(B$-$V) in mag and the reference for the observed V$-$band magnitudes are, respectively: SN 1979C: 43970, 16.14, 0.18, \protect\citep{Barbon1982a,DeVaucouleurs1981a,Ferrarese2000}; SN 1998S: 50874.7, 17.0 , 0.22 \protect\citep{Fassia2000}; ASASSN$-$15nx: 57218.6, 127.5, 0.07 \protect\citep{Bose2018}; SN 2013fc: 56516.2, 83.2, 0.94, \protect\citep{Kangas2016}; SN 1980K: 44540.5, 5.6, 0.3, \protect\citep{Barbon1982b}; SN 2014G: 56669.6, 24.5, 0.21 \protect\citep{Terreran2016}; SN 2008es: 54574, 1008, 0.012, \protect\citep{GezariDISCOVERY2008es}; SN 1987A: 49849.8, 0.05, 0.16, \protect\citep{Hamuy1990}.}
\label{fig:lc_comparison}
\end{figure*}

The lightcurve of SN~2016gsd shown in Fig. \ref{fig:lc} displays a slightly rounded peak, followed by a linear decline across all wavelengths. At brightest, SN~2016gsd has an absolute magnitude of $V=-19.95\pm0.08$ mag, (including the uncertainty in z) and then fades at a rate of 0.033$\pm~0.001$ mag per rest frame day, with the uncertainty derived from a linear regression. We can not say whether the SN is still rising at the onset of our LCO monitoring, but we do see the peak within the time period sparsely covered by ATLAS. Thus, we take the peak as the midpoint of the latest ATLAS observation where the SN is rising and the earliest LCO observation in which the SN is falling, i.e. MJD~$= 57662_{-5}^{+5}$. The decline in bluer filters is faster: the \textit{B}$-$band lightcurve fades by 0.058$\pm~0.002$ mag per rest frame day in the +20 to +101 rest frame day range. The NIR bands are not particularly well sampled, but show a slightly shallower decline than at optical wavelengths. This is consistent with an overall reddening of the spectral energy distribution over time.

After +101 rest frame days our data becomes sparse. The V band decline rate from +101 to +123 rest frame days, inferred from the next observation only, is faster, at 0.05 $\pm~0.01$ mag per rest frame day. For the next two points, the +123 to +167 rest frame day period, the decline rate has changed to 0.023 $\pm~0.005$ mag per rest frame day with uncertainty derived from the error bars on the photometry. The decline rate observed is significantly faster than that expected from full trapping of gamma rays from the decay of $^{56}$Co \citep[0.0098 mag per day;][]{Woosley1989}. Our latest detection is at 286 rest frame days in r band, and the decline rate inferred from +167 to +286  rest frame days would be 0.008$\pm~0.002$ mag per rest frame day, implying that the decline was significantly slower during this period. However, the last r band magnitude may suffer from some host galaxy contamination, as we could not successfully perform template subtraction for this image. Therefore, this late time decline rate may be underestimated.


With an absolute magnitude at peak of $V=-19.95\pm0.08$ mag, SN~2016gsd is among the brightest type II SNe ever seen (with the notable exception of super$-$luminous type II SNe such as SN 2008es \citep{GezariDISCOVERY2008es} or CSS121015:004244+132827 \citep{Bene14}). In Fig. \ref{fig:lc_comparison} we show SN~2016gsd with a set of similarly bright type IIL SNe, including the prototypical SN 1979C \citep{Barbon1982b}, along with ASASSN$-$15nx \citep{Bose2018}, SN 2013fc \citep{Kangas2016} and SN 1998S \citep{Fassia2000}. All of these events have a peak V band absolute magnitude of $\sim-$20 mag, and show a linear decline. Notably, SN 1998S was originally classified as a type IIn SN, although the narrow lines in the spectrum disappear by a few weeks after the discovery \citep{Fassia2000}. For comparison, we show in Fig. \ref{fig:lc_comparison} the ``normal'' type IIL SNe 1980K and 2014G which demonstrate the gap in luminosity between these objects and the ``bright'' type IIL events such as SN~2016gsd. With the exception of ASASSN$-$15nx, the SNe brighter than $V=-$19 all show evidence for interaction, which could explain their increased luminosity as compared to type IIP and "normal" type IIL SNe. We also show SN 2008es \cite{GezariDISCOVERY2008es}, as an example of a type II SN that is much more luminous still than the rest of the SNe shown, although with similar spectral evolution (see Sect. \ref{sect:spec}). It was suggested that SN 2008es was also powered by strong CSM interaction \citep{Miller2009,Bhirombhakdi2019}.

\begin{figure}
\includegraphics[width=0.5\textwidth]{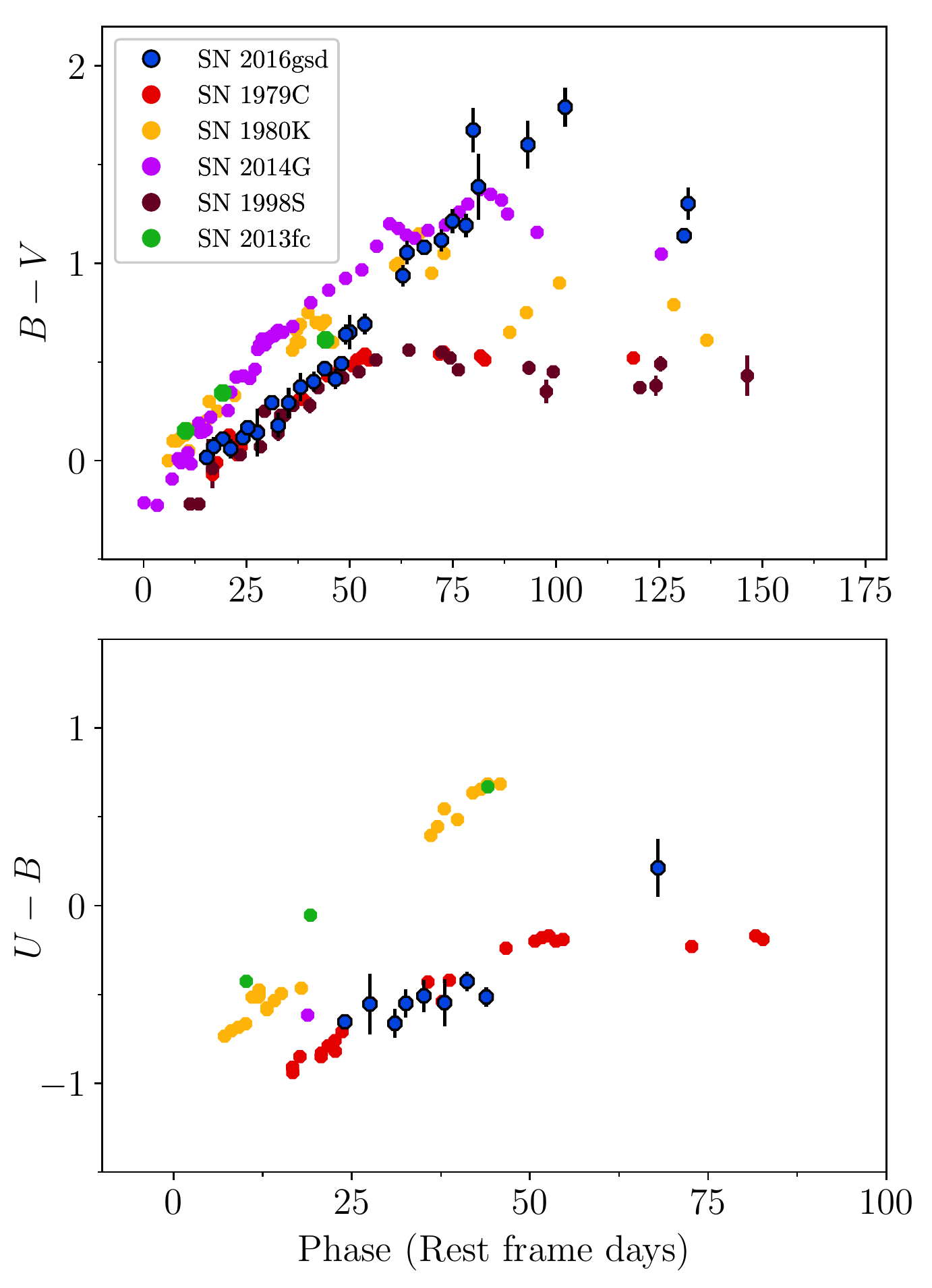}
\caption{The colour evolution of SN~2016gsd compared to other type II SNe. References and correction parameters used are as in Fig. \ref{fig:lc_comparison}}
\label{fig:colour_comp}
\end{figure}

In Fig. \ref{fig:colour_comp} we show the $B$-$V$ and $U$-$B$ colour evolution of SN~2016gsd, compared to the set of comparison type II SNe shown in Fig. \ref{fig:lc_comparison}. In this respect SN~2016gsd appears to show a broadly similar colour evolution to other type II SNe. From +10$-$100 rest frame days, the SN becomes steadily redder as the photosphere cools; and a similar colour change is seen in the comparison SNe. This seems to continue slightly longer than typical for this group of objects and in the final points SN~2016gsd becomes slightly bluer again. For the first 50 days the evolution is quite similar to that of SN 1998S and SN 1979C but after this point those objects transition to a constant colour. In $U$-$B$, SN~2016gsd is rather similar to SN 1979C, and is significantly bluer than the less luminous type IILs. We note that the $U$-$B$ colour for SN~2016gsd is already very blue compared to other type II events with only the Galactic reddening taken into account. Therefore we assume that there is little to no additional reddening in the host galaxy of the SN.

We attempted to observe SN~2016gsd at late times in NIR K band but it was not detected. We derive an upper limit shown in Tab. \ref{tab:JHK_phot}. Other type IIL SNe such as SN 1979C and SN 1998S have shown an excess of NIR luminosity in K band at late times, which has been interpreted as an echo from circumstellar dust \citep{Dwek1983,Mattila2001,Gerardy2002,Pozzo2004}. Our upper limit in K band is not deep enough to constrain SN~2016gsd from having a similar excess to these objects.

\subsection{Explosion epoch}
\label{subsec:Explosion epoch}
\begin{figure}
\includegraphics[width=0.5\textwidth]{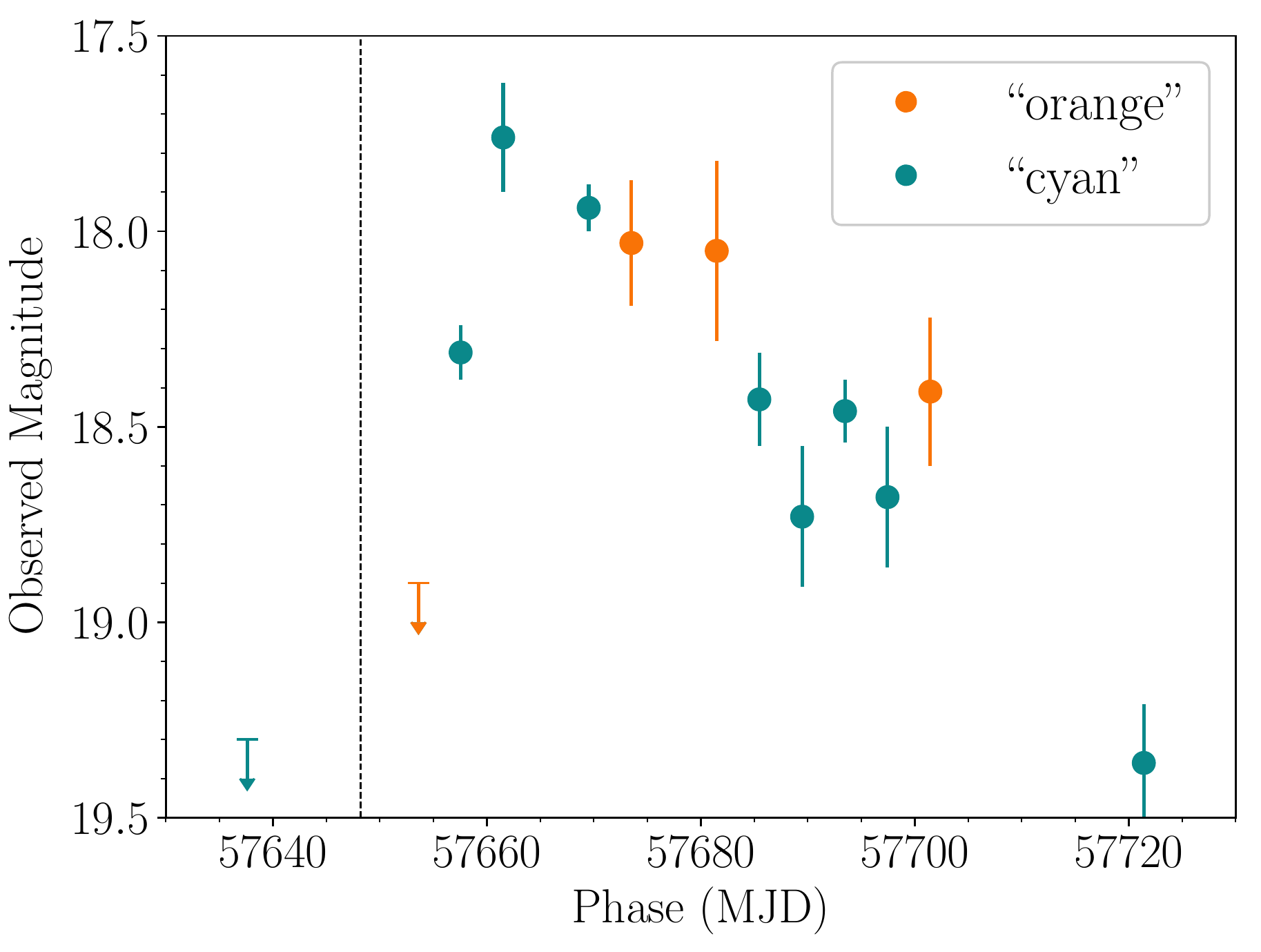}
\caption{The ATLAS lightcurve for SN~2016gsd, including the pre$-$discovery non$-$detections that are used to constrain the explosion epoch. The dashed line shows the adopted explosion date.}
\label{fig:ATLAS_lc}
\end{figure}

The most constraining data for determining the explosion epoch of SN~2016gsd are the ATLAS pre$-$discovery images listed in Table \ref{tab:ATLAS_phot}, and shown in Fig. \ref{fig:ATLAS_lc}. Our earliest detection is on MJD 57657.55, at $c=18.31\pm0.07$ mag, which is 3 days before the discovery by K. Itagaki on MJD 57660.80. The field of SN~2016gsd was last observed by ATLAS in the \textit{c} filter around 20 days prior, so this is not particularly informative. However, we do have a non$-$detection in the \textit{o} filter four days earlier, when the SN was not detected to a limiting magnitude of $o>18.9$. Based on the earliest $B$-$V$ colour we have for SN~2016gsd, we infer a blackbody temperature of 20,000 K, which is consistent with what is seen in young type II SNe. For this temperature, using the synthetic photometry code {\sc synphot} we expect an ATLAS $c-o$ colour of $\sim-$0.3 mag. If this was the colour of SN~2016gsd soon after explosion, the non$-$detection on MJD 57653.61 in the {\it o} filter implies that the {\it c} filter magnitude at this time must have been fainter than $c=18.6$ and that the SN brightened by more than 0.3 mag in the four days before our first detection.

\begin{table}
\centering
\begin{tabular}{|c|c|c|c|c|} \hline
Phase \\ (days)&Date (UT) &MJD&c&o\\ \hline
$-$10.4 & 2016$-$09$-$06 & 57637.6    & $>$19.3       & $-$        \\
5.6 & 2016$-$09$-$22 & 57653.6    & $-$             & $>$18.9  \\
9.6 & 2016$-$09$-$26 & 57657.6	& 18.31 $ \pm $ 0.07  & $-$        \\
13.5 & 2016$-$09$-$30 & 57661.5	& 17.76 $ \pm $ 0.14  & $-$        \\
21.5 & 2016$-$10$-$08 & 57669.5	& 17.94 $ \pm $ 0.06  & $-$        \\
25.5 & 2016$-$10$-$12 & 57673.5	& $-$             & 18.03 $ \pm $ 0.16	\\
33.5 & 2016$-$10$-$20 & 57681.5	& $-$             & 18.05 $ \pm $ 0.23  \\
37.5 & 2016$-$10$-$24 & 57685.5	& 18.43 $ \pm $ 0.12  & $-$        \\
41.5 & 2016$-$10$-$28 & 57689.5	& 18.73 $ \pm $ 0.18  & $-$        \\
45.5 & 2016$-$11$-$01 & 57693.5	& 18.46 $ \pm $ 0.08  & $-$        \\
49.5 & 2016$-$11$-$05 & 57697.5	& 18.68 $ \pm $ 0.18  & $-$        \\
53.5 & 2016$-$11$-$09 & 57701.5	& $-$             & 18.4 $ \pm $ 0.19 \\
73.4 & 2016$-$11$-$29 & 57721.4	& 19.36 $ \pm $ 0.15  & $-$        \\
\hline
\end{tabular}
\caption{ATLAS {\it c} and {\it o}$-$band photometry for SN~2016gsd in the AB system. Phases are relative to the explosion epoch of MJD~$= 57648_{-4}^{+8}$ obtained in Sect. \ref{subsec:Explosion epoch}.
\label{tab:ATLAS_phot}}
\end{table}

As the ATLAS data are not particularly deep, our pre$-$discovery limits do not strongly constrain the explosion epoch. However, the ATLAS data do show a rise in {\it c} filter brightness between the first and second detection, suggesting that the SN was discovered relatively soon after explosion. As the explosion epoch for SN~2016gsd we adopt the midpoint of the last non-detection and the first detection with the {\it c} filter,  MJD~$= 57648_{-10}^{+10}$ (2016 Sept. 17 UT). The uncertainty is the range between these two values. Given that the peak was on MJD$=57662_{-5}^{+5}$ as discussed above, the rise time would then be 14 days. This would be consistent with the average rise time of 13.6$\pm$0.6d observed for type IIL SNe by \citet{Gall2015} and with the sample of \citet{GG2015} who found an average rise time of 10.6$_{-4.2}^{+7.9}$ d from their sample of 48 type II SNe.

\subsection{Pseudo$-$bolometric luminosity}
\label{sect:bol}

\begin{figure}
\includegraphics[width=0.5\textwidth]{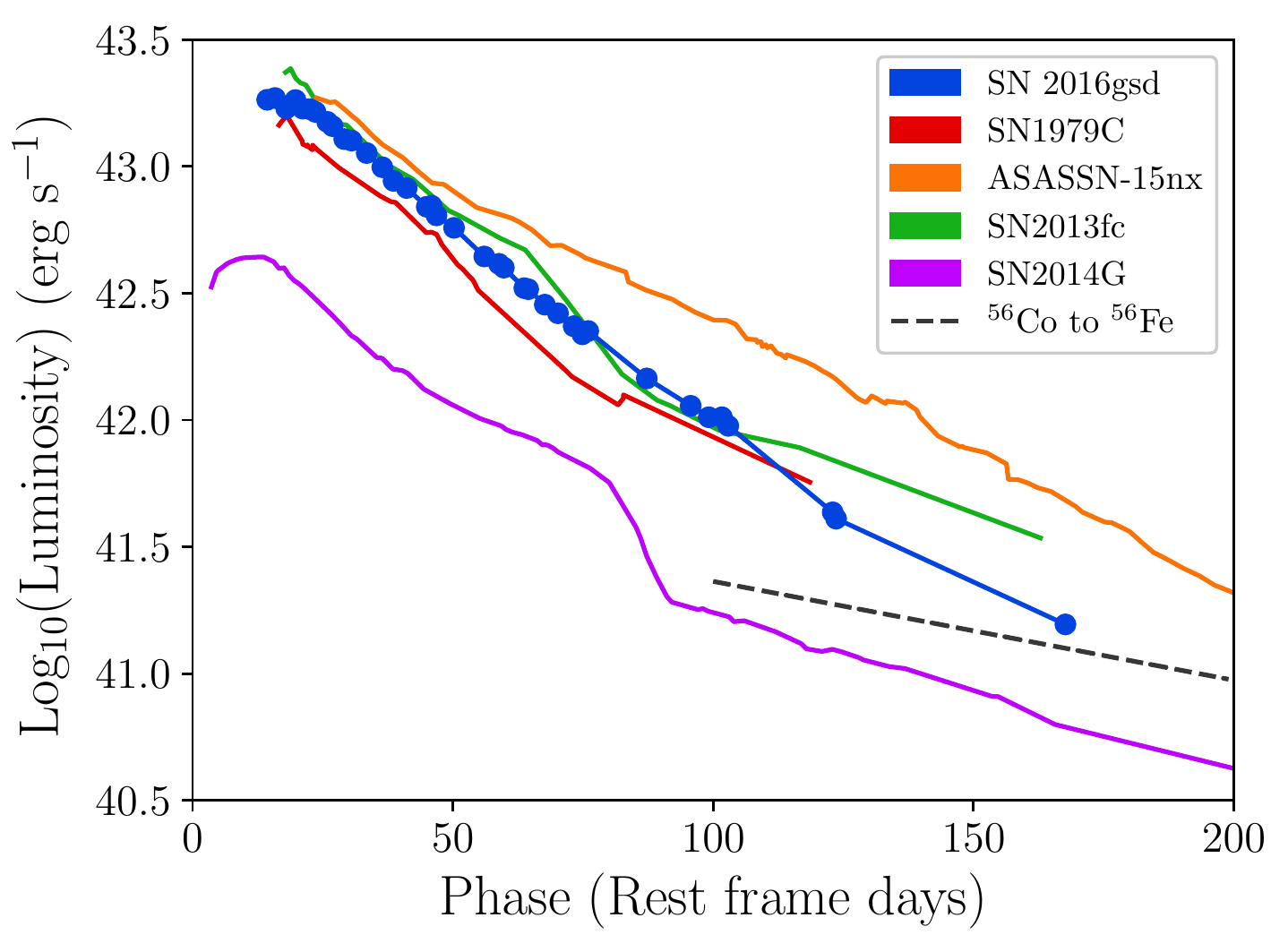}
\caption{The optical pseudo$-$bolometric luminosity in the 3000\AA~$-$ 8500\AA~range for SN~2016gsd, compared to other SNe. Values calculated for SN 1979C, SN 2013fc and SN 2016gsd, other values taken from literature with the sources given in the caption of Fig. \ref{fig:lc_comparison}.  Adopted parameters and references are the same as in Fig. \ref{fig:lc_comparison}.}
\label{fig:bol}
\end{figure}

We estimated the optical pseudo$-$bolometric luminosity of SN~2016gsd by integrating under the flux in the 3000\AA~$-$ 8500\AA~range, i.e. \textit{UBgVri} filters, interpolating where data was unavailable in a particular band. We chose not to include the \textit{JHK} data, as considerable interpolation and extrapolation would have been required given the small number of points. The fluxes were derived via conversion from the broad band filter magnitudes, corrected for distance and extinction. The resulting pseudo$-$bolometric luminosities are shown in Fig. \ref{fig:bol}, along with a selection of comparison objects. Estimating the SN rise as a simple line, the resulting total radiated energy over the measured pseudo$-$bolometric light curve (0 to +167 rest frame days) provides a lower limit for the total radiated energy of $6.5\times10^{49}$ erg. 

Although we omit the NIR data from the full analysis, we can measure the luminosity in the 1.165$\mu$m$-$2.282$\mu$m range as we did for the optical luminosity, at the epochs where we have these data. At $\sim$50 rest frame days, the NIR luminosity was $\sim$25\% of the pseudo-bolometric luminosity. At $\sim$100 rest frame days, this value was $\sim$60\%. If we extrapolate the NIR luminosity evolution, we would expect the NIR luminosity to be equivalent to the optical at $\sim$125 rest frame days. 

While we do not observe SN~2016gsd settling onto a clearly radioactively powered tail phase, we can still use the bolometric luminosity to set an upper limit to the ejected $^{56}$Ni mass. Comparing to the bolometric luminosity of SN 1987A \citep{bol87A}, for which we assume an ejected $^{56}$Ni mass of 0.075 M$_\odot$, we find $M(^{56}\mathrm{Ni})<0.12$ M$_\odot$ for SN~2016gsd, based on the bolometric luminosity at +167 rest frame days. This method assumes similar gamma ray trapping occurs in SN~2016gsd as that which occurred in SN 1987A at the measured point. The slope implied by our sparse data in the +131$-$167 rest frame day period is steeper than the expected decline rate for the $^{56}${Co} powered phase, which could imply that the trapping is less efficient and therefore that we have a higher mass of $^{56}$Ni than these limits. However, if the luminosity we observe at this time is due to CSM interaction (as discussed in Section \ref{sect:disc2}), we can not infer this.

\subsection{Blackbody fitting}
\label{sect:BB}

\begin{figure}
\includegraphics[width=0.48\textwidth]{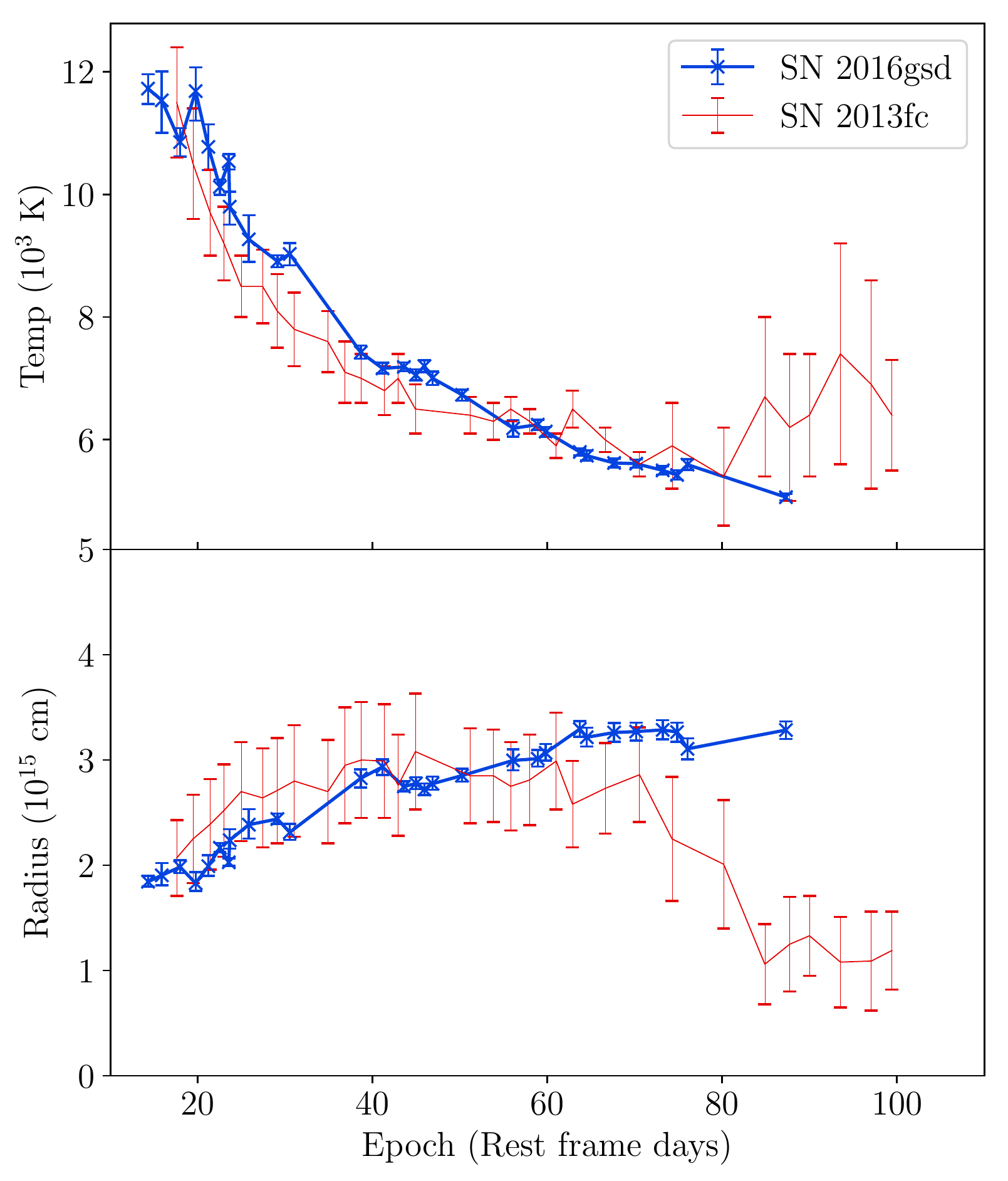}
\includegraphics[width=0.48\textwidth]{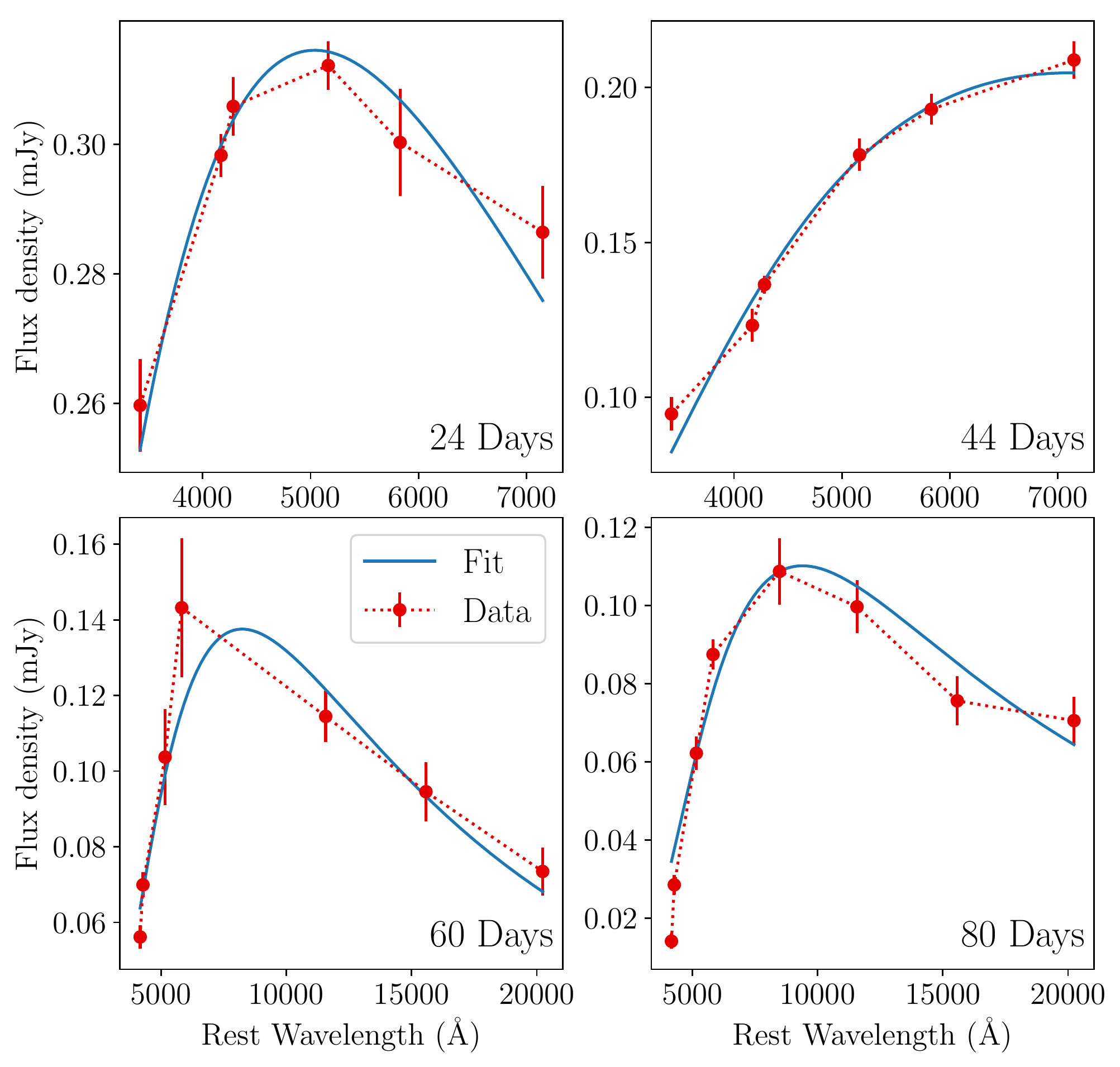}

\caption{Upper panel: The evolution of the temperature and radius of the fitted blackbody to the photometric data of SN~2016gsd. Points represent the median and errorbars represent the 16th and 84th percentile of the parameters of the MCMC samples. Lower panel: Examples of the blackbody fit to the observed spectral energy distributions.}
\label{fig:BB}
\end{figure}

Using the fluxes derived from our photometry, we fit blackbody functions to the SN SED during its evolution. We fit the available photometric bands, linearly interpolating where data was unavailable in a particular band. We did not interpolate over the \textit{z} band, as the data are very sparse. We omitted the \textit{U} band after +44d, as the data becomes too sparse to reliably interpolate. We also omitted the \textit{i} band from our fits after +44d, as at this redshift the H$\alpha$ emission line falls in this band, and the spectra from this epoch show considerable H$\alpha$ emission. We included the NIR \textit{JHK} bands in the +57$-$115d period where they are available. We found that the optical data alone could not give a good fit in the period of 44$-$57d, leading to a temperature discontinuity when we begin fitting the NIR data. We therefore extrapolated the NIR points for the +44$-$57d period, which gave a continuous temperature evolution. Thus, we fit \textit{UBgVri} from +15$-$44d, \textit{BgVrJHK} from +44$-$67d and \textit{BgVrJHK} from 67d to 105d. 

\begin{figure*}
\includegraphics[width=1\textwidth]{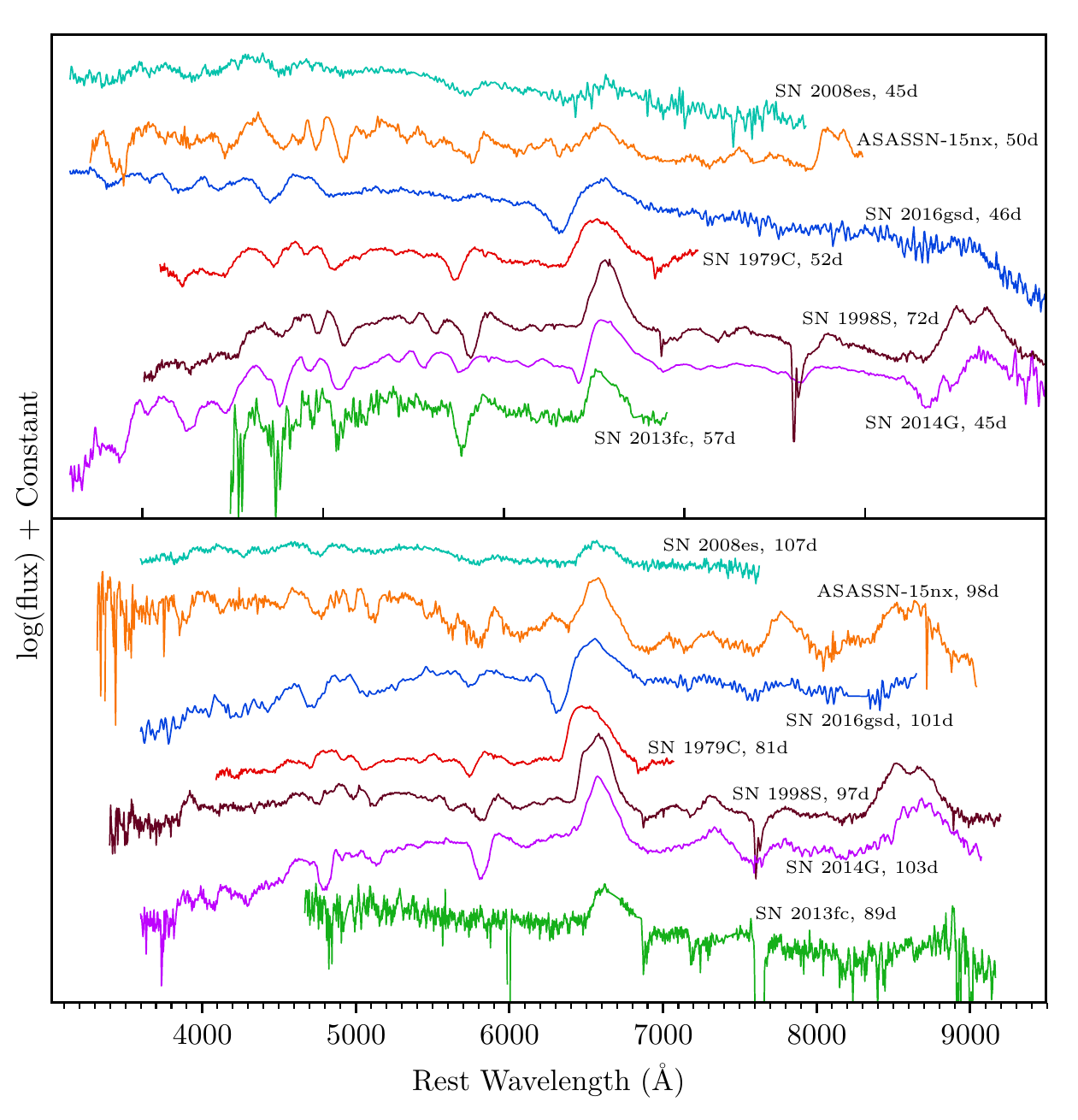}
\caption{Comparison of SN~2016gsd with the type II SNe SN~2008es \citep{Miller2009}, ASASSN$-$15nx \citep{Bose2018}, SN 1979C \citep{Barbon1982b}, SN 2014G \citep{Terreran2016}, SN 1998S \citep{Fassia2000} and SN 2013fc \citep{Kangas2016} at close to +50 (top) and +100 days (bottom). All spectra are shown in the rest wavelength. Spectra have been offset for visual clarity.}
\label{fig:spec_comp}
\end{figure*}

To perform the fitting, we used the EMCEE python implementation \citep{Foreman2013} of the Markov chain Monte Carlo (MCMC) method to estimate the best$-$fit blackbody parameters. The evolution of the temperature and the radius of the fitted blackbody is shown in Fig. \ref{fig:BB}, compared to SN 2013fc \citep{Kangas2016}, which is one of the SNe that had a similar \textit{V} band light curve and colour evolution (shown in Fig. \ref{fig:lc_comparison} and Fig. \ref{fig:colour_comp}). We also show our data and the fitted blackbody to demonstrate the quality of the fit at some selected epochs. These comparisons, along with comparisons to the optical spectra, confirmed that the fits were reasonable up to $\sim$ 90 rest frame days, although as the photospheric emission becomes less important at late epochs the fits are expected to be worse as the blackbody approximation becomes less appropriate. The evolution of the blackbody parameters is similar to SN 2013fc, which is to be expected as the light curve and colour evolution is fairly similar from +20$-$60d. To fit SN 2013fc, \cite{Kangas2016} required an additional $\sim$2000K blackbody component from +75d as their optical blackbody underestimated the required NIR flux. We do not find evidence this is needed. 

Using the values for temperature and radius derived from the fits, we can calculate the luminosity from the Stefan$-$Boltzmann law, and by integrating under the resulting light curve (as in \ref{sect:bol}) we measure the total radiated energy of the SN. This value for the 0$-$93 rest frame day period is $1.3 \times 10^{50}$ erg, as compared to $6.2 \times 10^{49}$ erg for our pseudo$-$bolometric LC over the same time period. The former value is approximately twice as large as the latter, which is unsurprising as our pseudo$-$bolometric results omit both the UV and NIR regions, which contribute significantly to the luminosity at early and late times respectively.

\section{Spectroscopic analysis}
\label{sect:spec}

A log of the spectroscopic observations is given in Tab. \ref{tab:Spectral log}. The spectral sequence is presented in Fig. \ref{fig:spec_seq}; the spectra are shown in their rest frame and have been corrected for extinction. Fig. \ref{fig:spec_comp} shows the spectra of SN~2016gsd along with spectra of a group of comparison objects at $\sim$50d and $\sim$100d.

The early spectra display a blue continuum, and while the precise temperature is uncertain as we do not cover the peak of the blackbody function, it is certainly greater than 10,000 K. In the +23 and +25 day spectra we see the Balmer sequence of H lines from H$\alpha$ as far as H$\epsilon$. H$\alpha$ has barely discernible absorption, while it is not clear in the higher order lines. Along with H, we see a broad P$-$Cygni line at 5876\AA\  which we associate with He~{\sc i}. The line profile of this line is extremely similar to the H$\alpha$ line at +23 and +25 days, but has become narrower by +34 days. Along with He~{\sc i} $\lambda$5876, we see a feature we associate with a He~{\sc i} transition at $\lambda$3889. This feature is strong in the +23 and +25 day spectra. Taking the case B recombination line ratios from \cite{Osterbrock1989}, He~{\sc i} $\lambda$3889 is almost as strong as He~{\sc i} $\lambda$5876, which is indeed what we observe in our data. The H Balmer transition at $\lambda$3889 is also present here, but is not strong enough to explain the feature we observe, compared to the other Balmer lines present in the spectrum. The He~{\sc i} $\lambda$3889 feature is weaker at +34d and gone by +44d. This is expected, as the decreasing temperature of the ejecta means we have less excitation of He~{\sc i} \citep{Dessart2010,Roy2011}. Here however, the He~{\sc i} lines persist considerably later than the $\sim$20 days seen in most type II SNe, reflecting that SN 2016gsd remains hotter for longer than usual.


The features observed in these first two spectra, taken about a week after the maximum light, are weaker than typically observed in type II SNe, and resemble the ``blue and featureless'' continua observed at very early times in type IIP SNe. This is also observed in SNe 1979C, 1998S and 2013fc. The model of \citet{Dessart2016} reproduces this effect by modelling the explosion of a star with a dense nearby shell of CSM generated by a wind of 0.1 M$_{\odot}$ per year that lasted a few years before the explosion. In this model, the simulated spectra at +20d and +34d show a blue continuum with only weak emission and shallow blueshifted absorption features. This is a good match with what is observed in SN 1998S and is attributed to the emitting region being at the edge of a hot, fast moving dense shell which has a steep density profile, causing shallow absorption and weak emission. 

\begin{figure}
\includegraphics[width=0.5\textwidth]{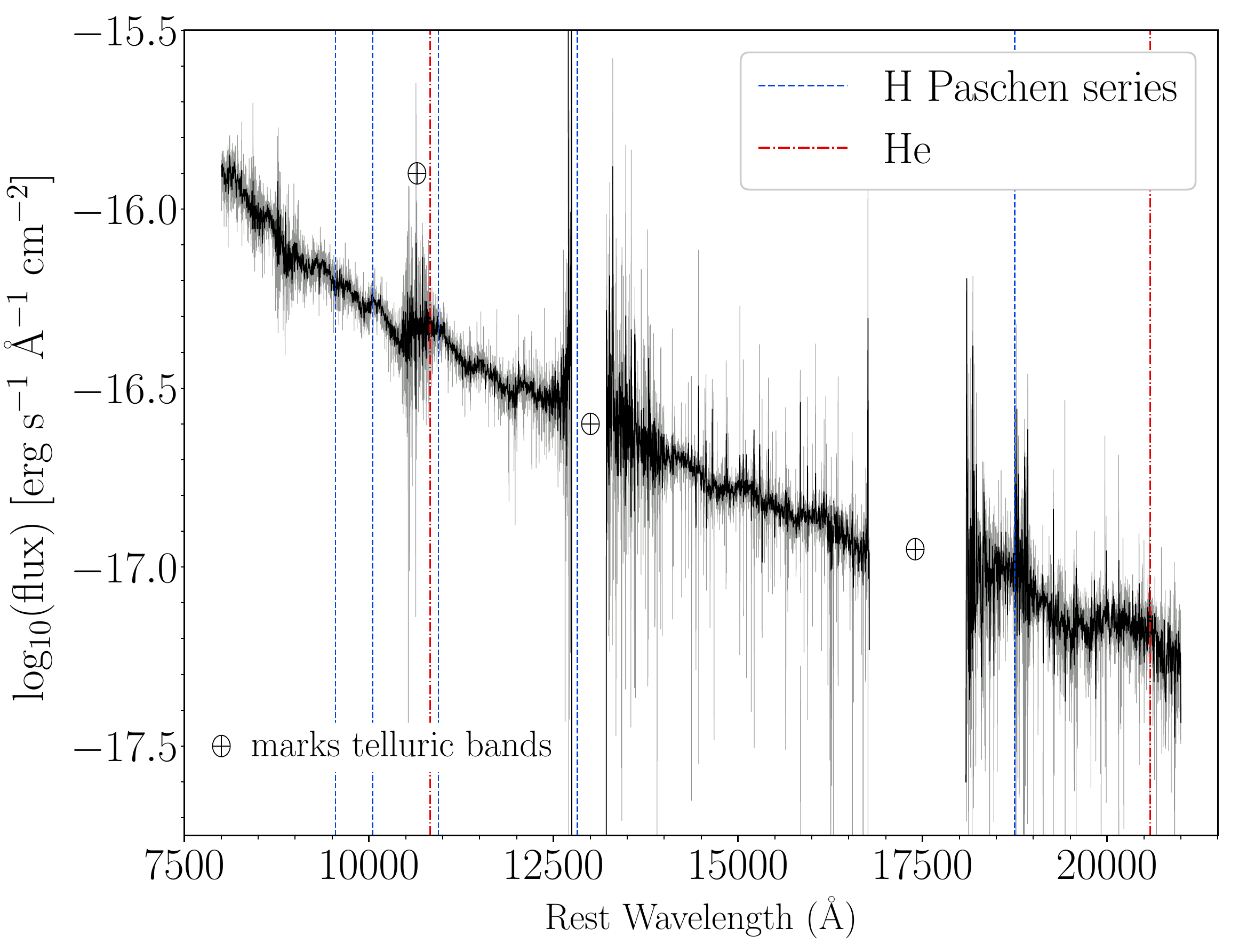}
\caption{NIR spectrum of SN~2016gsd.
The rest frame wavelengths of the lines of the Paschen series and He~{\sc i} are marked. The shown spectrum is heavily smoothed, with the unsmoothed spectrum shown in the background in grey}
\label{fig:NIR_spec}
\end{figure}

The NIR spectrum of SN~2016gsd taken at +25 days is shown in Fig. \ref{fig:NIR_spec}. The spectrum has low S/N, with no obvious strong features. We identify a possible broad, shallow P$-$Cygni line around 1.08$\upmu$m, although the feature is unclear due to the nearby telluric band. We identify the feture with He~{\sc i}$\lambda$10830, given the presence of He~{\sc i}$\lambda$5876 in the contemporaneous optical spectrum. The NIR spectrum can be matched smoothly with the optical spectrum, with no sign of a NIR excess or large deviation from a blackbody spectral energy distribution.

The +34 days spectrum of SN~2016gsd is redder, and the Balmer lines in the blue, including H$\beta$, have become more prominent. At +45 days, we see the emergence of features around H$\beta$ associated with metal species. We see a slightly raised continuum redwards of H$\beta$, formed from multiple Fe emission features, but the broad Fe absorptions seen in type II SNe at this time are not well defined. We see evidence for  emission and absorption features typically associated with metal species in type II SNe emerging around the H$\beta$ feature. The lines associated with He~{\sc i} other than $\lambda$5876 are gone in this spectrum, with the $\lambda$3889 line going from prominent to completely absent. It seems likely then that the line at $\lambda$5876 is now predominantly Na~{\sc i}~D, and that it should arise at the same time as the other metal lines is predicted by, for example, \citet{Dessart2010}. The Na~{\sc i}~D line is also much weaker than in other type II SNe at this time. There is no evidence for the presence of the Ca~{\sc ii} NIR triplet despite it being common at this point in type II SNe. \cite{Gutierrez2017a} found only one object (SN~2009aj) in their sample of 122 type IIs that did not show this feature after +25 days, and noted that it was an object that showed evidence of CSM interaction at early times.

At +68 days, the typical absorption lines are stronger, but still weak compared to other type II SNe at +44 days. At this point there begins to be evidence for an intermediate velocity line within the emission feature of H$\alpha$ with FWHM $\sim$ 2700~\kms. This becomes more clear in the later spectra, and particularly in the +132 day spectrum (see Sect. \ref{subsect:vels}). We discuss the H$\alpha$ line profile further in Sect. \ref{subsect:vels}. At +102 days and then +132 days, the Na~{\sc i}~D absorption begins to be more prominent. At +68 and +102 days, the Ca~{\sc ii} NIR triplet appears to become visible, although the feature is at the edge of the CCD. By +132 days the Ca~{\sc ii} NIR triplet is clear and a broad feature becomes visible at $\sim$7250\AA. This feature is most likely identified as the [Ca~{\sc ii}] $\lambda$7291,7324 doublet but is blueshifted from the expected position by 1750 \kms. This feature is weak or not at all present in earlier spectra, in contrast with many of the objects shown in Fig. \ref{fig:spec_comp}.

\begin{figure}
\includegraphics[width=0.5\textwidth]{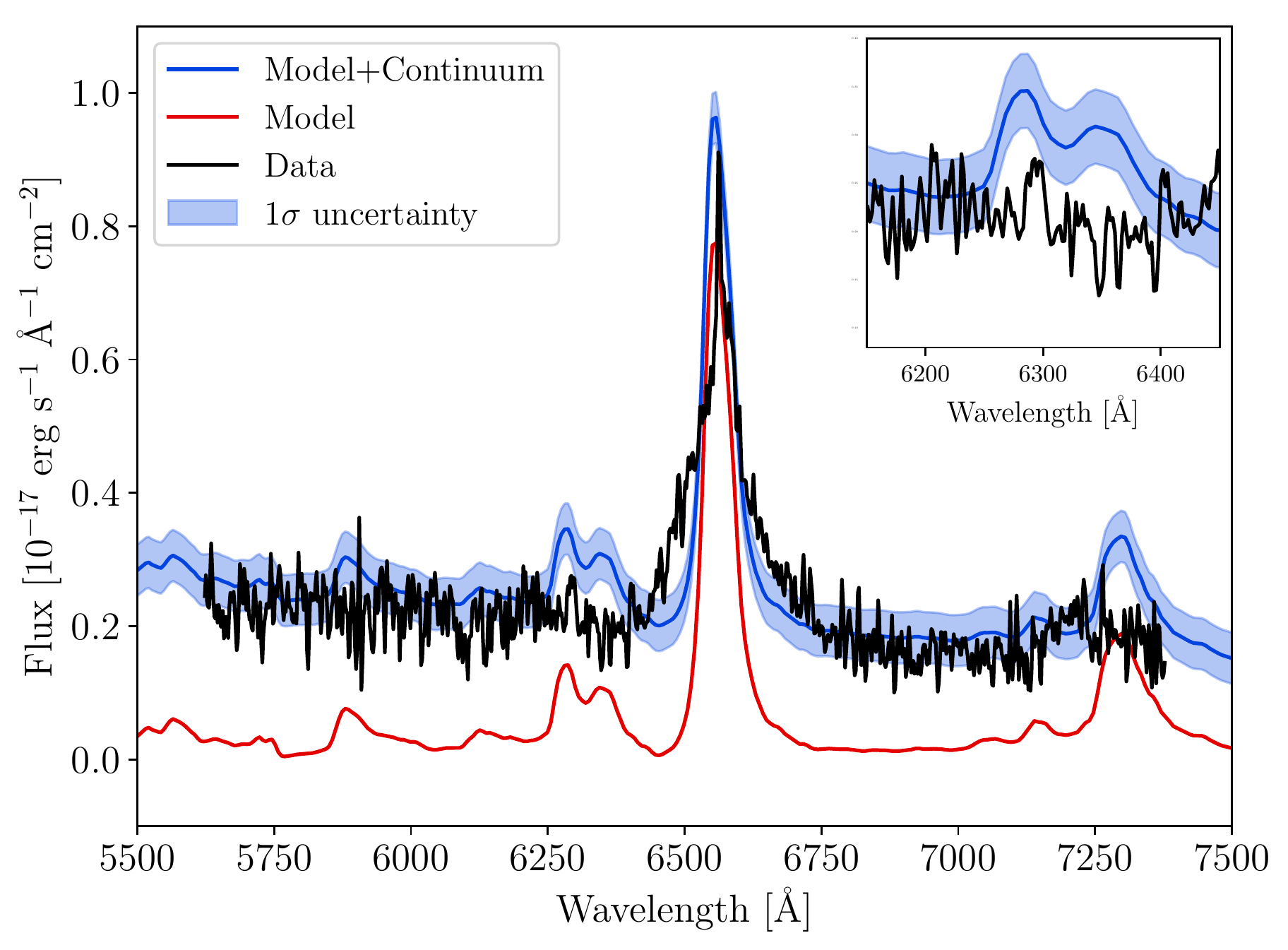}
\includegraphics[width=0.5\textwidth]{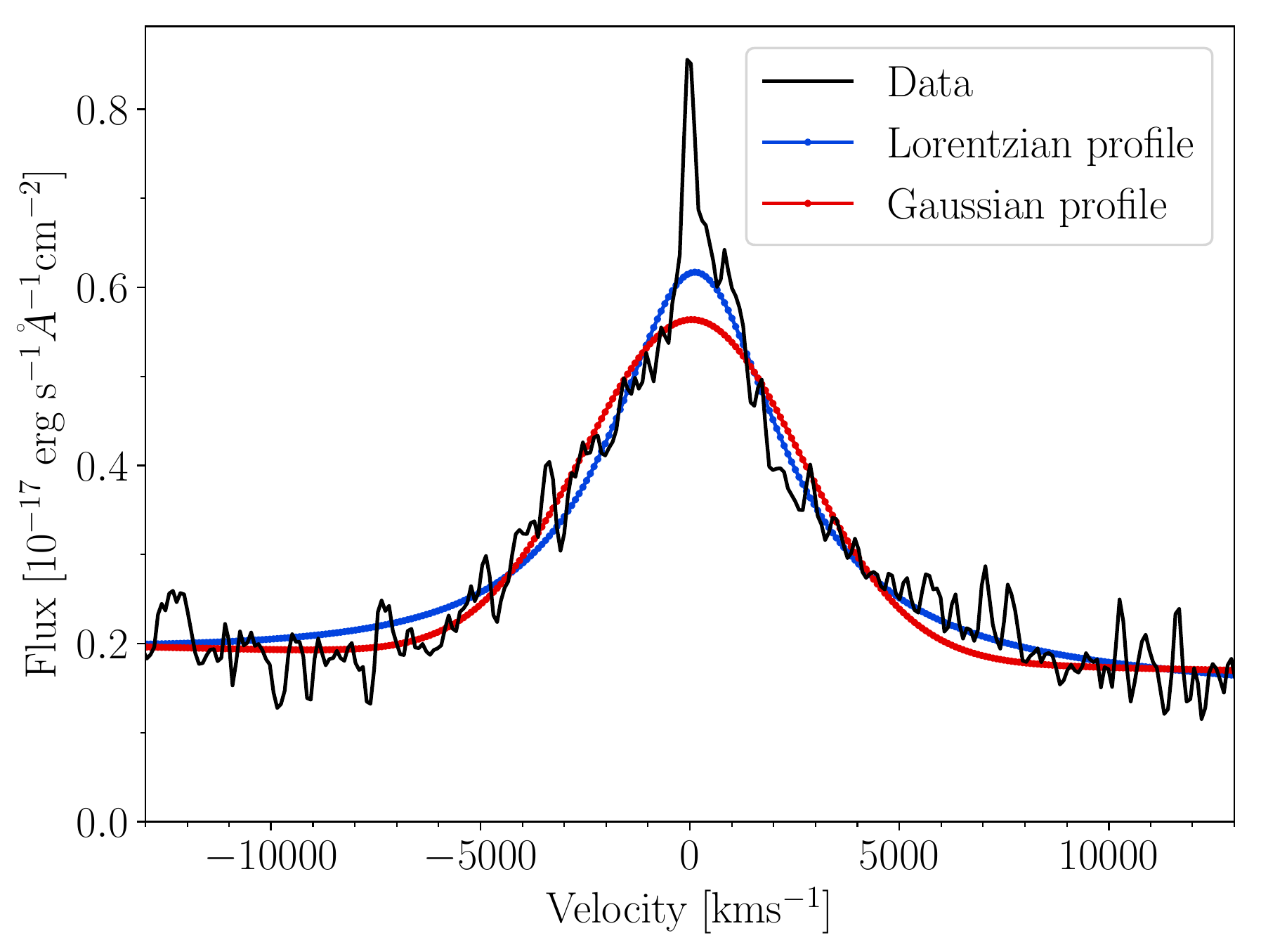}
\caption{Upper panel: The +301 rest frame day spectrum of SN~2016gsd, shown in black, as compared to a model of the nebular spectrum obtained for the explosion of a 15 M$_\odot$ ZAMS red supergiant at +306d, shown in red (\protect\citealp{Jerkstrand2014}). The model flux has been scaled to correspond to the distance of SN~2016gsd. In blue we show the model with the addition of the continuum found in the data. The shaded region shows the uncertainty derived from the data and applied to the model. The inset shows the region around the [O~{\sc i}]$\lambda$6300,6364 emission lines. Lower panel: H$\alpha$ line profile in velocity space, with the  best fit broad Lorentzian profile overplotted in blue, and the broad Gaussian profile in red.}
\label{fig:late_spec}
\end{figure}

One striking aspect of the spectral sequence is the weakness of the metal lines. Type II SNe typically have spectra that at bluer wavelengths are dominated by absorption by many metallic lines \citep[e.g.][]{Dessart2011}. In the case of SN~2016gsd, these lines are weak or absent, while the Na~{\sc i}~D absorption and the Ca~{\sc ii} NIR triplet are also atypically weak. Weaker metal lines in type II SNe may be explained with a low metallicity progenitor, both by models \cite[e.g][]{Dess13} and with reference to observations \cite[e.g][]{Taddia2016,Gutierrez2018}. Such a progenitor is also consistent with the host of SN~2016gsd, which appears to be a faint dwarf galaxy (Sect. \ref{sect:Host}). Furthermore, the metal lines in SN 2016gsd are somewhat weaker than in SN 1998S and SN 1979C (which both come from bright spiral galaxies) at similar times (see Fig. \ref{fig:spec_comp}). However, the ionisation conditions in the ejecta also have a large effect on these lines \cite[see e.g.][]{Utrobin2005,Dessart2008}.

Most type IIL SNe have weaker absorption compared to emission in H$\alpha$, whereas SN~2016gsd shows strong absorption throughout.  A simple measurement of this is the ratio of the equivalent widths of the absorption and emission parts of the line profile. For SN~2016gsd this ratio is between $\sim$0.4 and 0.6 in the +34d to +69d period. \cite{Gutierrez2014} suggested that type II SNe with stronger absorption relative to the emission in H$\alpha$ display lower velocities; whereas SN~2016gsd shows consistently fast ejecta (Sect. \ref{subsect:vels}). Explanations for the weak H$\alpha$ absorption in many type IIL SNe include extra emission (perhaps associated with CSM outside the photosphere) that ``fills in'' the absorption, or a low envelope mass or steep density gradient in the envelope \citep{Schlegel1996OnSupernovae,Gutierrez2014}.
For SN~2016gsd, the  presence of a strong absorption may suggest that the envelope is not low mass, although this would have strong implications for the shape of the light curve. We address this further in Sect. \ref{sect:Model}.

Our final spectrum obtained at +301 rest frame days shows only a broad emission feature of H$\alpha$ with FWHM $\sim$ 5000~\kms. We do not clearly see any of the typical forbidden lines such as [O~{\sc i}]$\lambda$6300,6364 or [Ca~{\sc ii}]$\lambda$7291,7323 that are seen in the nebular spectra of core$-$collapse SNe. In the upper panel of Fig. \ref{fig:late_spec}, we compare the observed spectrum to the model for the explosion of a 15 M$_\odot$ red supergiant with an ejected $^{56}$Ni mass of 0.062 M$_\odot$ at +306 rest frame days from \cite{Jerkstrand2014}. The $^{56}$Ni mass of the model is well below our upper limit of $M(^{56}\mathrm{Ni})<0.12$ M$_\odot$.
The observed spectrum was scaled so that synthetic photometry in {\it r}$-$band matched the observed {\it r}$-$band magnitude at +285 rest frame days. This photometric measurement was not based on image subtraction, so the brightness of the SN may be overestimated due to host galaxy contamination. However, if the photometry is contaminated by the host, the continuum level in the spectrum would be lowered, and lines would be relatively stronger, making it more difficult to conceal them. We linearly fit the continuum of the observed spectrum and add the result to the model. We made a measurement of the standard deviation in the 5750\AA$-$6250\AA~region of the observed spectrum, and the shaded region shows the uncertainty associated with this measurement, applied to the model spectrum.

We do not clearly see the presence of, [O~{\sc i}] $\lambda$6300,6364 or [Ca~{\sc ii}]$\lambda$7291,7323 in the spectrum of SN~2016gsd. There is evidence for a feature close to the [Ca~{\sc ii}]$\lambda$7291,7323 position but it is significantly blueshifted from the expected position, although not enough to be consistent with [Fe~{\sc i}]$\lambda$7155. These lines are both clear in the model spectrum. A lower $^{56}$Ni mass could naturally weaken these lines, while higher than usual densities could still suppress the forbidden lines at this phase.

In the lower panel of Fig. \ref{fig:late_spec}, we show the line in velocity space, as compared to both a Lorentzian profile with a FWHM of $~$5000 \kms and a Gaussian profile with a FWHM of $~$6000 \kms. The Lorentzian profile fits the red edge of the line and more of the peak well, while the Gaussian profile fits the blue edge better while significantly underfitting the peak and red edge. A Lorentzian profile is typically associated with scattering of photons in an optically thick CSM. However this requires a dense scattering medium, which is the opposite of what we expect at late times in an expanding SN, unless there was ongoing interaction with dense CSM. We discuss this scenario in Sect. \ref{sect:disc2}).

We note that there is also a narrow component visible on top of the broad line fits. Simultaneously fitting a Lorentzian/Gaussian to the broad component and Gaussian to the narrow line we find it consistent with an unresolved narrow line at the instrumental resolution corresponding to 295 \kms. This could be related to the SN, and would indicate the existence of un$-$shocked CSM around the SN, or be associated with the host galaxy. We do not see any other narrow lines from the host galaxy in any of our observations of SN~2016gsd. In our spectrum of the host galaxy we see an H$\alpha$ emission line that is $\sim$20 times stronger in flux than this narrow feature. Although this spectrum is of a different region than the SN, we can use this to determine that the narrow line is most likely associated with the host. We discuss the late time SN spectrum and its implications for the progenitor of SN~2016gsd further in Sect. \ref{sect:Discussion}.


%


\begin{figure}
\includegraphics[width=0.5\textwidth]{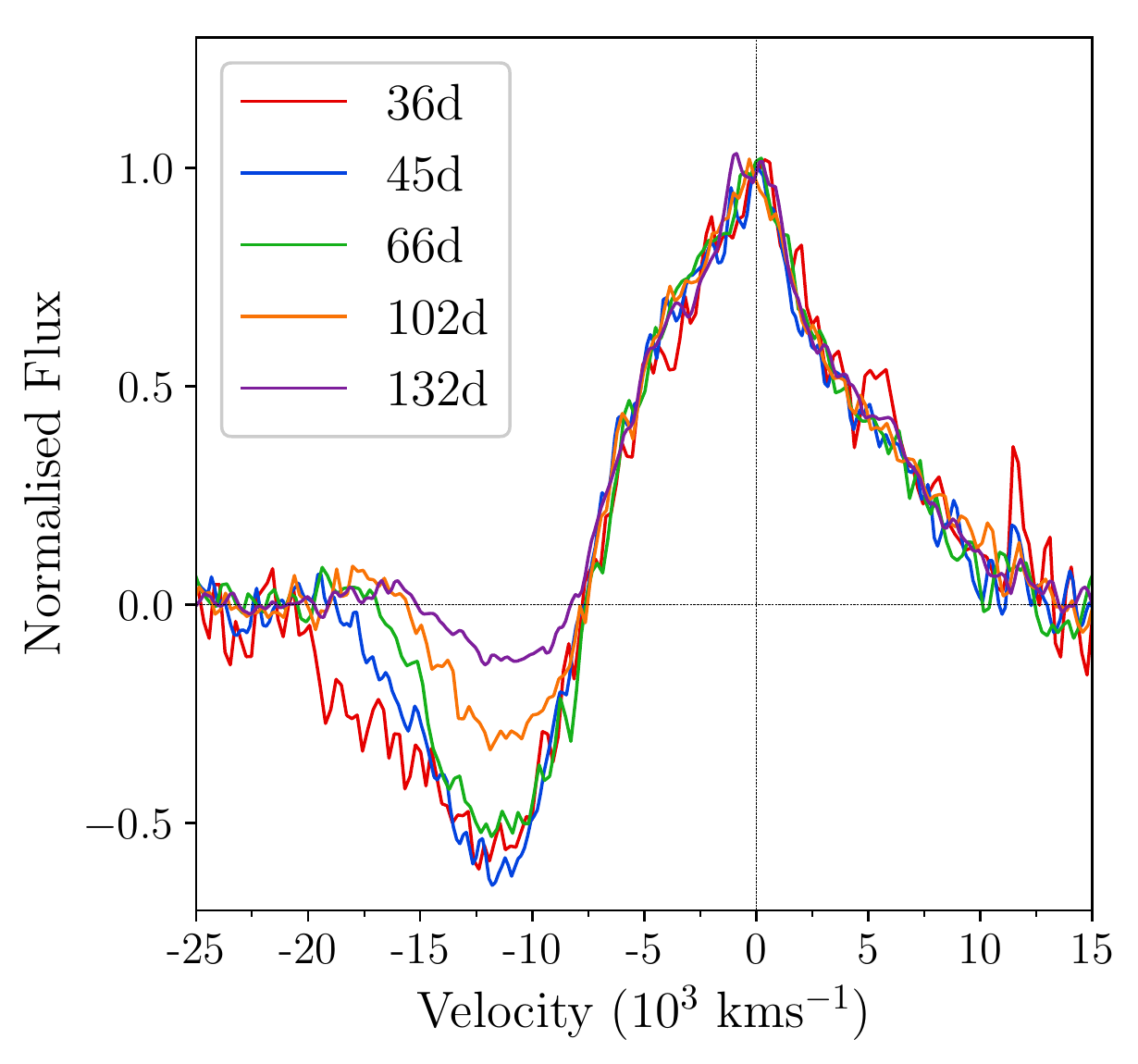}
\caption{The evolution of the line profile around H$\alpha$ in velocity space. The spectra have been normalised: first the continuum around the line was subtracted, and then the spectra was scaled such that the peak of the emission is at 1.}
\label{fig:lineprofile}
\end{figure}

\subsection{Velocities}
\label{subsect:vels}

\begin{figure}
\includegraphics[width=0.5\textwidth]{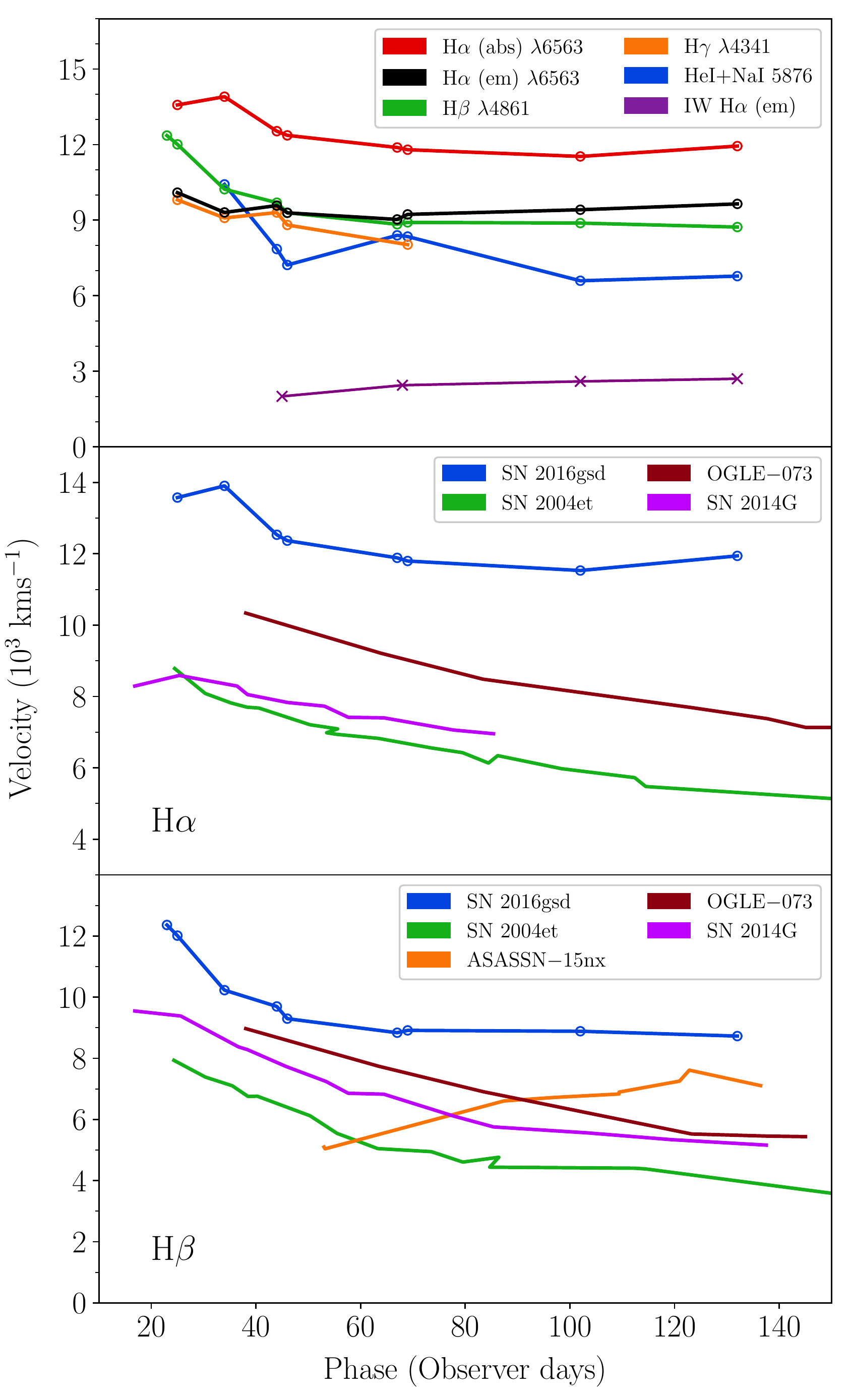}
\caption{Velocity evolution of prominent spectral features of SN~2016gsd, and comparisons with other type II SNe. Except where noted otherwise, velocities were measured from the minimum of the absorption and no attempt was made to deblend the P$-$Cygni profile. However, the width of the intermediate width H$\alpha$ (IW H$\alpha$) emission in the top panel is adopted from line profile fitting such as that shown in Fig. \ref{fig:multicompHalpha}. Values for OGLE$-$2014$-$SN$-$073 were taken from \citet{Terreran2017}, all other values were taken from the sources listed in the caption of Fig. \ref{fig:spec_comp}.}
\label{fig:velocities}
\end{figure} 

We measured line velocities from the centre of the absorption for all clear spectral features by fitting a Gaussian to the absorption line component while simultaneously fitting a straight line to the local background to approximate the continuum. No attempts were made in the measurements to simultaneously fit both components of the P$-$Cygni profile.

The velocity derived from the absorption in H$\alpha$ is high, with an initial velocity of $\gtrsim$14000 \kms, that subsequently declines to around 12000 \kms. In Fig. \ref{fig:lineprofile} we show the evolution of the flux in the H$\alpha$ region. The blue edge of the absorption line, tracing the fastest moving gas, is initially at $\sim$20000 \kms~ and declines approximately linearly to $\sim$15000 \kms~ through the $\sim$100 days of observation. H$\beta$ shows a similar evolution to H$\alpha$, although $\sim$2000 \kms~ slower. These velocities are much higher than that seen in H lines in typical type II SNe \cite[e.g.][]{Gutierrez2017}; this can be seen from Fig. \ref{fig:velocities} where we compare SN~2016gsd to a number of type II SNe, including the ``normal'' type IIP SN 2004et \citep{Maguire2010}, type IIL SN 2014G \citep{Terreran2016} and OGLE$-$2014$-$SN$-$073 \citep{Terreran2017}, which displays some of the highest photospheric phase velocities measured in a type II SN. Velocities measured from H$\alpha$ and H$\beta$ absorption features in IILs have been observed to be larger than in IIP supernovae, particularly after +50d.

Very high velocities similar to SN~2016gsd (up to $\sim$20000 \kms) are observed in H absorption in the early spectra of SN~2000cb \citep[see][]{Utrobin2011}. These high velocities require a higher explosion energy, but likely less than 2 x 10$^{51}$ erg \citep{Dessart2019a}. However, this was the explosion of a compact blue supergiant progenitor similar to the case of SN 1987A and very different from the case of SN~2016gsd.

We also measure velocities derived from the feature at $\sim$ 5900\AA. In the first 3 spectra at 23d, 25d and 34d, we assume this feature is predominately He~{\sc i} $\lambda$5876, from our identifications of other lines of He~{\sc i} described above. After this, we assume the line has become dominated by Na~{\sc i}~D. This is consistent with other metal lines appearing at this time, implying ionisation conditions are favourable for Na~{\sc i}~D absorption features to appear. There is no clear absorption in the first 2 spectra. The first absorption we can measure gives a velocity of $\sim$10200 \kms~for He~{\sc i}, which decreases and settles at  $\sim$7500 \kms.

\begin{figure}
\includegraphics[width=0.5\textwidth]{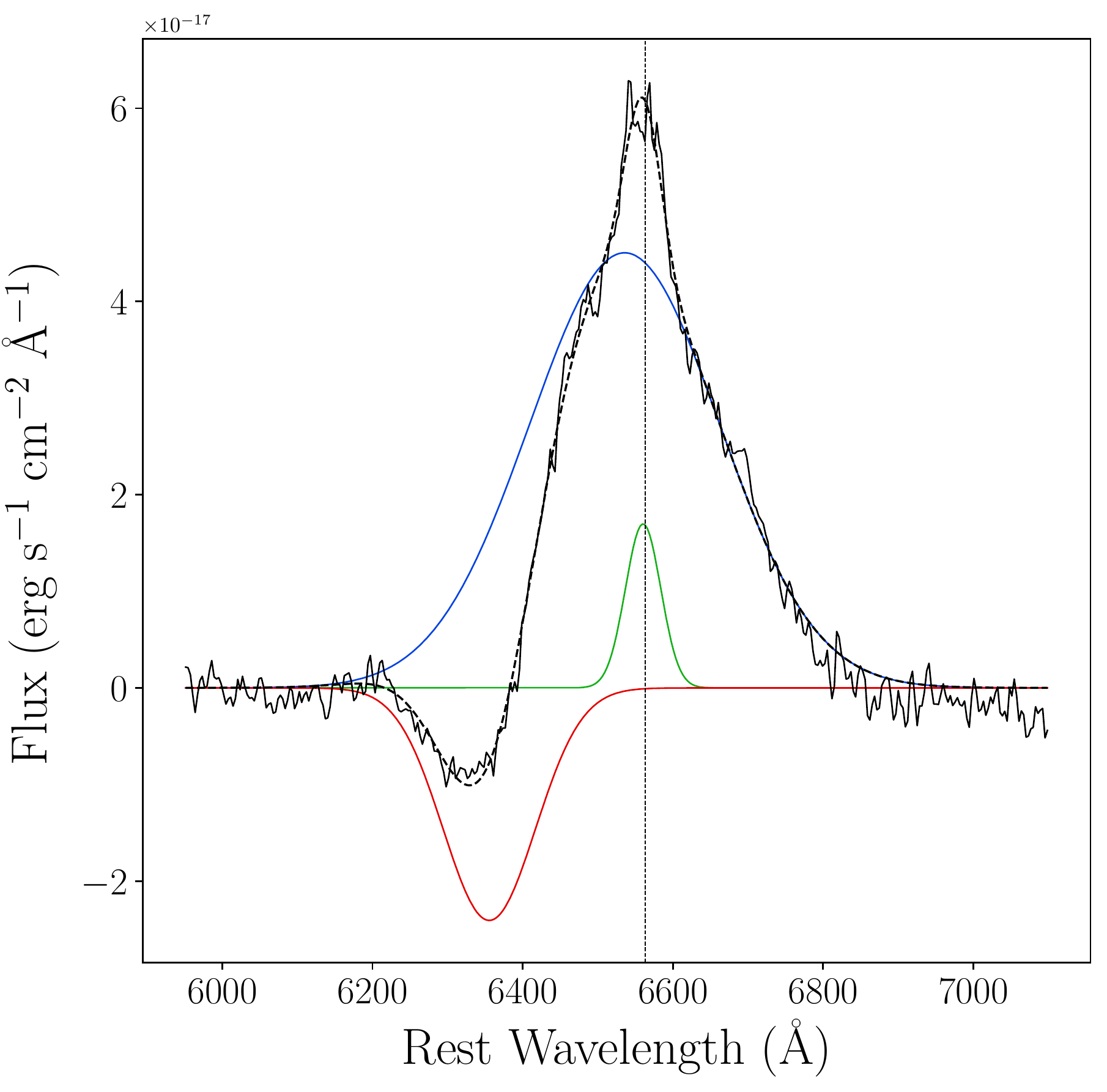}
\caption{The H$\alpha$ profile +132d described by a combination of three Gaussian profiles, representing a P-Cygni profile as well as an intermediate width feature with FWHM $\sim$2700 \kms~at the rest wavelength of H$\alpha$. The broad H$\alpha$ emission is blueshifted by $\sim$1100 \kms. A linear fit to the continuum has been subtracted from the data.}
\label{fig:multicompHalpha}
\end{figure}

In the period from +36d to +132d the peak of the broad H$\alpha$ emission feature is fixed at the rest wavelength (see Fig. \ref{fig:lineprofile}). This is unlike the vast majority of type II SNe: in \citet{Anderson2014b} it is shown that a blueshifted emission peak is a generic feature seen in observations of type II SNe, and this is supported by modelling. The effect is due to the blocking of the redshifted emission from the far side of the ejecta by the optically thick photosphere, which is efficient due to the steep density profile within the ejecta \citep{Dessart2005a}. In the case of SN 2016gsd, one possible explanation for the lack of blueshift is the presence of a secondary emission component at the rest wavelength which is concealing the blueshift of the emission feature formed in the ejecta. This is shown clearly in Fig. \ref{fig:multicompHalpha}, where we simultaneously fit 3 Gaussian profiles to the H$\alpha$ feature. The line is well fit by the addition of a intermediate width feature with FWHM  $\sim$ 2700 \kms~ to the typical broad P-Cygni features. The intermediate width feature is at approximately the rest wavelength, while the broad emission is blueshifted. The spectra from 45d-132d can all be similarly fit with a blueshifted broad emission feature with line centre taking values from approximately $-$2500 to $-$1000\kms~ and an intermediate width feature that increases in width from approximately 2000 to 3000\kms, as shown in Fig. \ref{fig:velocities}. The blueshifts in the broad emission peaks we see in these fits are consistent with those measured in the sample of \citet{Anderson2014b}. Intermediate width emission features are commonly seen in type IIn SNe. We discuss this further in Sect. \ref{sect:disc2}.

\section{Host galaxy}
\label{sect:Host}

The host galaxy of SN~2016gsd has a visible nucleus and extended, asymmetrical, diffuse structure. We measured the magnitude of the galaxy from deep imaging of the SN site at +489 days, after the SN had faded. This was measured with aperture photometry, using an aperture centred at the brightest pixel and increased in size until we captured close to all the galaxy flux. The galaxy was previously catalogued by the Pan-STARRS 3Pi survey \citep{Chambers2016,Flewelling2016} as PSO J040.1438+19.28236. However the Kron radius used for these measurements is smaller than the apertures we use and only captures the flux of the bright nucleus, excluding the diffuse asymmetric flux of the host. Both sets of measurements are given in table \ref{tab:Hostphot}, compared with values for the Small Magellanic Cloud (SMC) and Large Magellanic Cloud (LMC). The host absolute magnitude is between the the absolute magnitude of the LMC and the SMC and somewhat redder than both objects in colour. It displays an offset nucleus and its maximum diameter, derived from the aperture we used to determine the magnitude, is $\sim$6.5'', corresponding to a projected size of $\sim$8.5 kpc at this redshift. The size of the Kron radius used in the Pan-STARRS $g-$band measurement was $\sim$4.5''.

\begin{figure}
\includegraphics[width=0.5\textwidth]{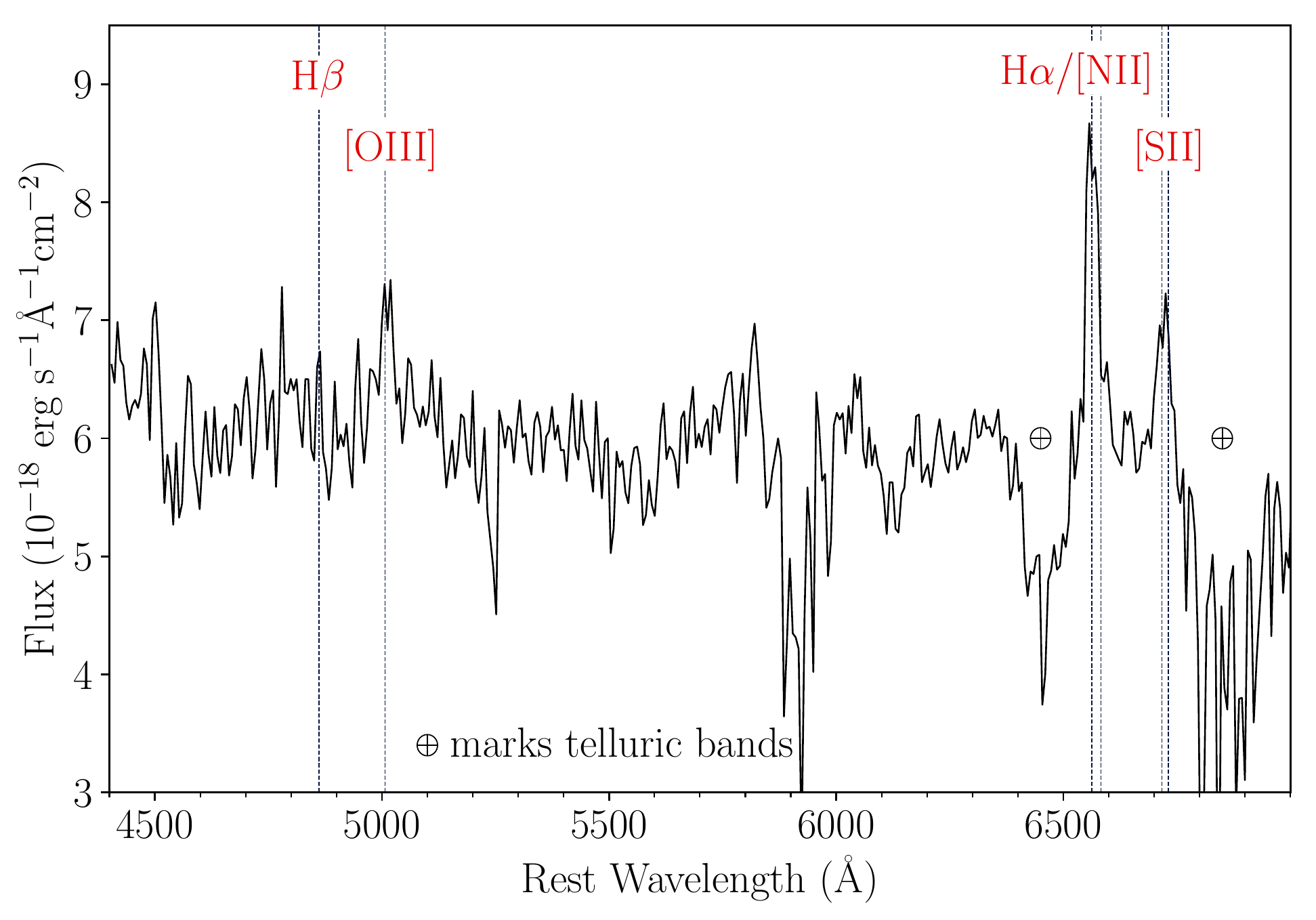}
\caption{Host galaxy spectrum. The lines used to estimate the metallicity are marked, along with [S~\textsc{ii}]. The feature at $\sim$5900\AA~ is the result of poor subtraction of sky emission lines and is not a physical absorption feature.}
\label{fig:Host_spectrum}
\end{figure}

Although we were unable to detect unambiguous host galaxy lines in any of the spectra taken of SN~2016gsd, we obtained a spectrum of the host (see Fig. \ref{fig:Host_spectrum}) in which we can detect emission features associated with [O~{\sc iii}]$\lambda$5007, [S~{\sc ii}]$\lambda$6717,6731 and a blend of [N~{\sc ii}]$\lambda$6583 and H$\alpha$. We do not detect an emission feature of H$\beta$. We set the redshift of the galaxy to z=0.067$\pm$0.001 based on fits to these emission features. We set upper limits for the flux ratios of:
\begin{equation}
\frac{[\textrm{O}~\textsc{iii}]\lambda5007}{\textrm{H}\beta} > 2.3 ; \quad
\frac{\textrm{H}\alpha}{[\textrm{N}~\textsc{ii}]\lambda6583} > 3.0 ,
\end{equation}
using spectrum statistics in the first case and simultaneously fitting the two blended features in the second. Following equations (1) and (2) in \citet{Marino2013}, we then find:
\begin{equation}
12 + \textrm{log(O/H)} < 8.35,
\end{equation}
which implies that the host of SN~2016gsd is at least as metal poor as the LMC, which has $12 + \textrm{log(O/H)} = 8.35$ \citep{Hunter2007}.

The host galaxy of SN~2016gsd is quite different from the hosts of the other bright type IIL supernovae such as SNe 1979C, 1998S and 2013fc that we compare to in this work. These SNe occurred in the spiral galaxies M100, NGC3877 and ESO 154-10 respectively, which are all 3-4 magnitudes more luminous than the host of SN~2016gsd in B-band (B-band magnitudes taken from \citet{vauc1991}, galactic reddening taken from \citet{Schl11} via the NASA Extragalactic Database (NED), distances are as listed in the caption of Fig. \ref{fig:lc_comparison} and taken from the sources therein). We can infer that these SNe likely came from higher metallicity environments.

\begin{table}
\centering
\begin{tabular}{|c|ccc|} \hline
&Host & SMC & LMC  \\ \hline
   B  & $-$17.10$^{a}$ $\pm$ 0.06 & $-$16.4 $\pm$ 0.1 & $-$17.85 $\pm$ 0.05 \\
   B$-$V & 0.77 $\pm$ 0.05 & 0.42 $\pm$ 0.03 & 0.44 $\pm$ 0.03 \\
\hline
   g & -17.23$^{b}$ $\pm$ 0.05 & $-$ & $-$ \\
\hline
\end{tabular}
\begin{flushleft}
$^{a}$ Photometry performed in deep imaging with NOT/ALFOSC \\
$^{b}$ Apparent magnitude taken from the Pan$-$STARRS 3Pi survey
\end{flushleft}
\caption{Host galaxy absolute magnitude and colour compared with values for the SMC and LMC taken from \citet{vauc1991}. All values are corrected for galactic extinction. Distances of 50~kpc \citep{Pietrzynski2019} and 62~kpc \citep{Graczyk2014} were adopted for the LMC and SMC respectively.
\label{tab:Hostphot}}
\end{table}

\section{Discussion}
\label{sect:Discussion}

\subsection{The observational class of bright type IIL SNe}
\label{sect:disc1}

SN~2016gsd is a member of a group of bright H$-$rich SNe that have a peak magnitude V<$-$19, and a long$-$lasting linear decline. This group includes SN 1979C \citep{Barbon1979}, ASASSN$-$15nx \citep{Bose2018} and SN 2013fc \citep{Kangas2016} which all showed broad Balmer lines consistent with a type II. However, there also are a number of SNe with a similar photometric evolution that show more clear signs of CSM interaction in their spectra, for example ASASSN$-$15no \citep{Benetti2018} and SN 1998S \citep{Fassia2000}. Furthermore, SNe such as SN 2008es \citep{GezariDISCOVERY2008es,Miller2009} have quite similar spectra to this group, but are much more luminous. It is hence unclear whether the group of bright, linearly declining type II SNe should be regarded as a single physical class, or whether their spectroscopic diversity points towards a number of distinct physical scenarios.

The peak brightness of all these objects could point towards a large radius for the photosphere at early times, requiring material at this distance from the progenitor star with appropriate mass and density. Such material could come about through a steady wind, eruptive mass loss events shortly before the SN explosion occurs or through binary interaction. These phenomena would create different configurations of material around the star, and could originate from stars with different initial parameters. The other important characteristic of this group, their linearity or lack of a plateau, is commonly explained by the density of the progenitor envelope being low. This prevents the cooling wave from creating a plateau, as it recedes too quickly through the envelope, and cannot maintain the luminosity of the early peak of the light curve. The low densities could be achieved, similarly to the peak brightness of the SN, through sufficient mass loss prior to the explosion.

There is evidence that CSM interaction also plays a role in the evolution of ``normal" luminosity type IIL SNe, i.e. those with peak magnitude <$\sim$-18.5. The type IIL SN 2014G \citep{Terreran2016} showed evidence for interaction through the weak/absent absorption in its H$\alpha$ line profile, as can be seen in Fig. \ref{fig:spec_comp}. Furthermore, boxy asymmetric emission features appeared around the H$\alpha$ emission after 100d. \citet{Bostroem2019} performed a detailed study of the nearby type IIL SN ASASSN-15oz to look for signs of interaction and find evidence for this from their radio observations and light curve modelling. They conclude that the progenitor underwent a short period of extreme mass loss shortly before the explosion.

From the theoretical side, there have been a number of attempts to systematically consider the effect of CSM on type IIP and IIL SN light curves with modelling. \citet{Morozova2017}  made use of the SuperNova Explosion Code (SNEC) to model three well studied type IIL SNe. They found that the addition of $\sim$0.1$-$0.5 M$_{\odot}$ of dense CSM with an extent of $\sim$1300$-$1900 R$_{\odot}$ improved the fit of their models. \citep{Hillier2019} discuss type II SN light curves and show that $\sim$1 M$_{\odot}$ of CSM extending from the progenitor surface to a few times 1000 R$_{\odot}$ can boost the peak luminosity to the level of ``normal" type IIL SNe such as SNe 2014G and 2013ej, while suppressing the measured velocities and reducing the absorption strength in the P-Cygni profiles. They comment that the exact configuration of the CSM (mass, extent and density structure) is important to reproduce the observed features. There are also further studies on the effect of CSM close to the progenitor which indicate that the early light curves of type II SNe are significantly affected by such material \citep[see e.g.][]{Morozova2018,Moriya2017}. However, SNe such as SN~2016gsd and SN~1979C are much more luminous than those discussed in these works, requiring a different configuration of CSM, if we assume the underlying progenitors and explosion mechanisms are not fundamentally different.

\subsection{SN~2016gsd as a luminous type IIL SN}
\label{sect:Model}

\begin{figure}
\includegraphics[width=0.5\textwidth]{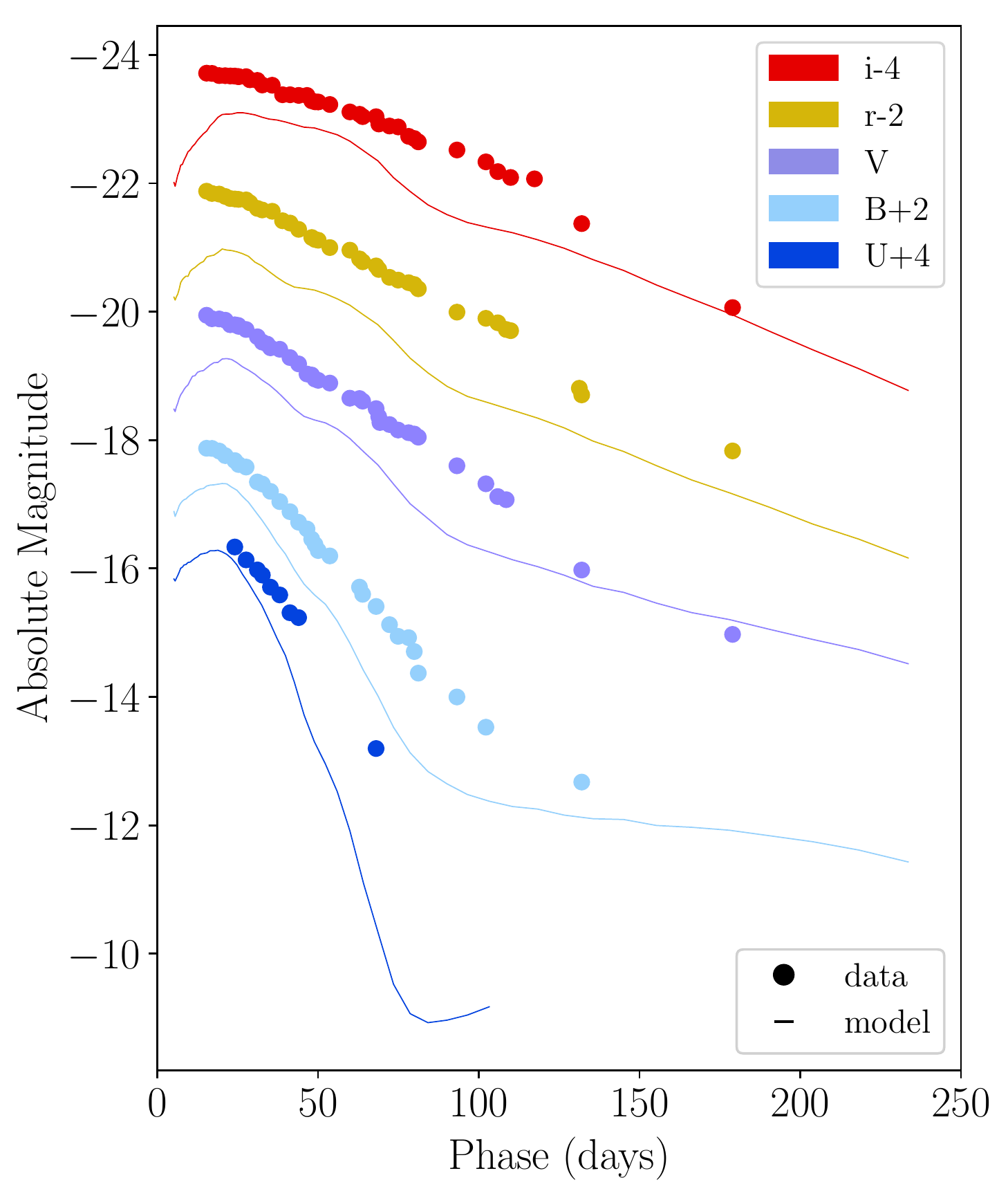}
\caption{Broad$-$band lightcurves of the {\sc jekyll} model compared to the observations of SN~2016gsd.}
\label{fig:jekyll_model}
\end{figure}

In \cite{Blinnikov1993}, the authors attempted to reproduce the peak brightness and linear decline of SN 1979C by exploding stellar models with relatively extended and low$-$mass H-rich envelopes and by taking into account the interaction of the SN with a presupernova wind. In these models, a large
peak
luminosity can be reproduced 
by 
the radiative energy in
a relatively 
extended envelope, whereas the tail luminosity can be reproduced by interaction of the ejecta with the presupernova wind. In addition, the low mass of the H-rich envelope gives a fast decline from peak luminosity, a characteristic of SN 1979C and other luminous type IIL SNe. The authors also explore a model (B7) with a high (0.2 M$_\odot$) mass of ejected $^{56}$Ni, in which case the tail luminosity is powered by the radioactive decay. Below we explore this case in more detail, using a similar model.

We use the NLTE spectral synthesis code {\sc jekyll} \citep{JEKYLL} to investigate if the observations of SN~2016gsd could be qualitatively reproduced by a model similar to model B7 by \citet{Blinnikov1993}, but without any contribution from CSM interaction. First we used {\sc mesa} \citep{Paxton2011, Paxton2013, Paxton2015, Paxton2018, Paxton2019} to construct a $M_{\mathrm{ZAMS}}=15$ M$_\odot$ pre$-$supernova model, artificially removed all but 2 M$_\odot$ of the H envelope and exploded the star with an explosion energy of $2 \times 10^{51}$ erg using the hydrodynamical code {\sc hyde} \citep{HYDE}.

\begin{figure}
\includegraphics[width=0.5\textwidth]{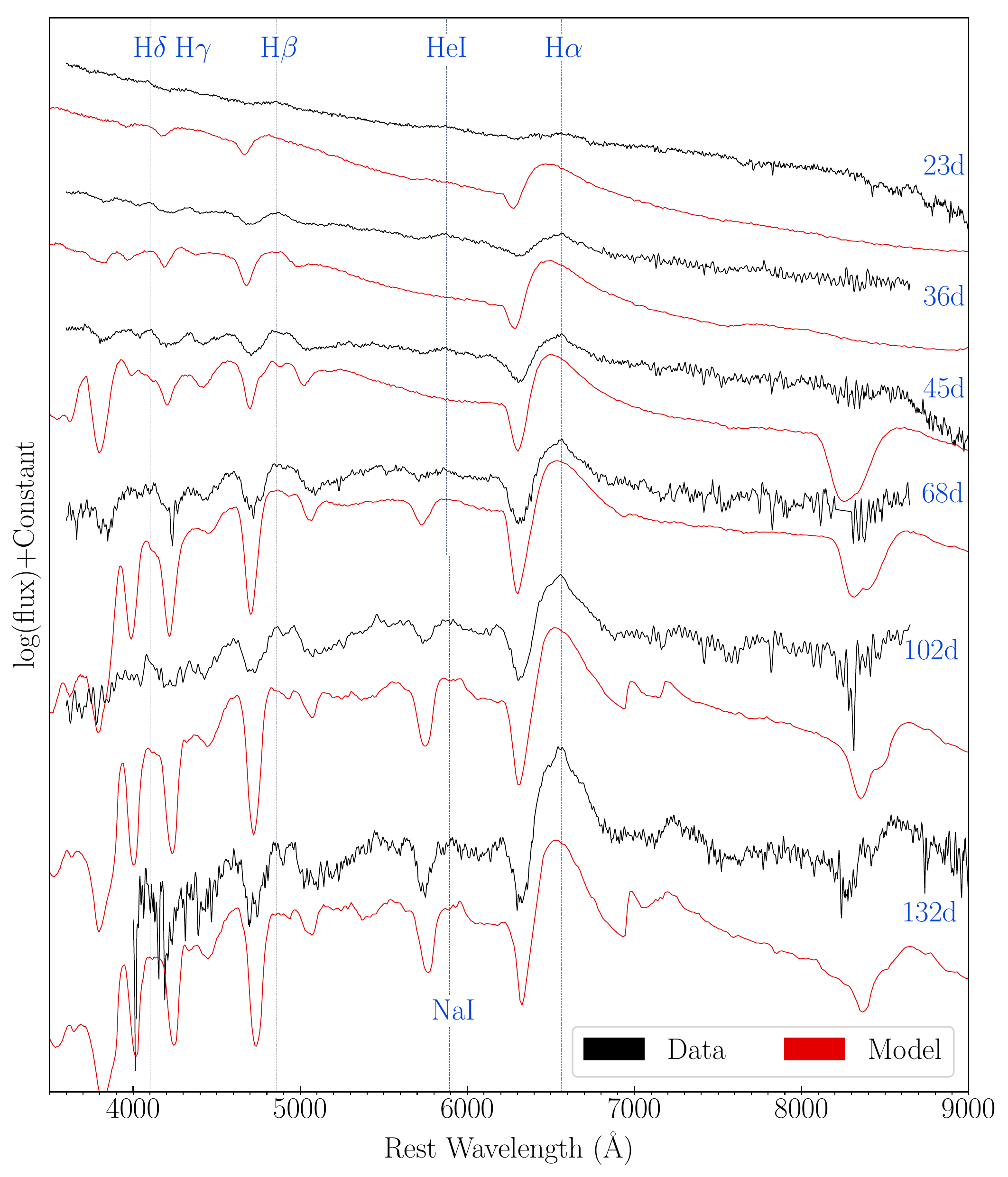}
\caption{Spectral evolution of the {\sc jekyll} model compared to the observations of SN~2016gsd.}
\label{model_spec}
\end{figure}
Explosive nucleosynthesis is not included, and the mass of ejected $^{56}$Ni was set to 0.2 M$_\odot$.
The {\sc hyde} model was then fed into {\sc jekyll} at +5 days and evolved to
+250 days.
However, before doing this we artificially increased the internal energy by a factor of four, to roughly mimic the effect of having a four times larger radius. This would correspond to a progenitor radius of 
3200 R$_\odot$.
Note, that this is about half the radius used in model B7, as the rise time otherwise becomes too long (see below). Otherwise, the {\sc jekyll} model is similar to model B7, but without an attached CSM.

Fig. \ref{fig:jekyll_model} shows the broad$-$band lightcurves and Fig. \ref{model_spec} the spectra of the {\sc jekyll} model compared to the observations of SN~2016gsd. 
The spectra is only shown for the first 100 days, as accurate modelling in the nebular phase requires a macroscopically mixed ejecta model \citep[see][]{JEKYLL}.
The model lightcurves are not as linear as SN~2016gsd and instead they show a short and weak plateau. The peak luminosity is 
about one magnitude lower and the quite long rise time of the model lightcurves 
seems to be in tension with the observations of SN~2016gsd.
Likewise, although the observed lines and their widths are similar, their strength is generally much weaker than in the model. The 
H$\alpha$ line is reasonably well fit from 46d onwards though, while the other Balmer lines and the metal lines are more prominent in the model than in our data throughout. The centres of the absorption in H$\alpha$, H$\beta$ and Na~{\sc i}~D in the model also match well with the data, indicating that the velocities are quite similar.

Although the model is promising in some respects, it certainly does not fully explain the observations of SN~2016gsd.
It does not reproduce the peak luminosity, the rise time, the linear decline or the strength of the spectral lines. Before considering what could be changed in the model, it should be noted that our assumption that increasing the internal energy is equivalent to inflating the radius is only an approximation, and that the parameters are based on model B7 rather than chosen to fit the SN~2016gsd data. Using a radius as large as that in model B7 would increase both the peak luminosity and the rise-time. Although this would better reproduce the peak luminosity, the rise time would become too long. Increasing the explosion energy in the model would also increase the peak luminosity, but would make the velocities too high. The relative strength of the metal lines in the model spectra compared to the data indicates that a lower-metallicity progenitor could improve the fit.

We can only speculate what is missing in the model, but it is likely that CSM interaction is needed to produce a better agreement (see earlier discussion in Sect. \ref{sect:spec}). Furthermore, 0.2 M$_{\odot}$ of $^{56}$Ni would cause the [O~{\sc i}] $\lambda$6300,6364 and [Ca~{\sc ii}]$\lambda$7291,7323 lines to be strong enough to be observable in the 320d spectrum of SN~2016gsd. This is evidence that these late time points are instead powered by interaction. 

\subsection{SN~2016gsd as an interaction powered event}
\label{sect:disc2}

Many bright, linear type II SNe discussed in this paper have clear indications of CSM interaction. SN 1979C was detected in the radio \citep{Lundqvist1988} and displayed a prominent NIR excess long after the SN had faded \citep{Dwek1983}. Analysis of SN 1998S inferred extensive interaction with two shells of CSM produced through steady winds or eruptions on different timescales and an overall high mass loss rate of $\sim$10$^{-4}$M$_{\odot}$yr$^{-1}$ \citep{Fransson1989,Pooley2002,Fransson2005,Mauerhan2012}.

\citet{Dessart2016} found that a model with $\sim$0.4 M$_{\odot}$ of extended CSM positioned close to the progenitor star was able to reproduce both the spectral characteristics and the light curve shape of SN 1998S for the first $\sim$80 days. The outer shell of CSM material was created by an intense period of mass loss (0.1 M$_{\odot}$ per year) that lasted only a few years. The material created in that period stretches out to $\sim$3000 R$_{\odot}$ with less dense material lying further out. This model reproduces well the spectral evolution of SN 1998S, where the spectra are blue with narrow lines at early times, blue with only shallow absorption features at $\sim$20$-$30d and finally evolve to show the broad emission and absorption features similar to type IIP SNe at $\sim$60$-$80d. The predominant remaining CSM signature at this point is the lack of absorption in the H$\alpha$ P-Cygni profile due to emission from the cool dense shell (CDS) that fills in this feature.  We consider whether a model similar to this would be suitable for SN~2016gsd and whether other CSM distributions would be appropriate.

Our first spectrum of SN~2016gsd was taken approximately 23d after the explosion and showed no narrow lines. If we assume that the fastest moving ejected material from SN~2016gsd was expanding spherically symmetrically at $\sim$ 20000 \kms~for this period, consistent with our measurements of the blue edge of H$\alpha$ absorption at +36d, the material swept up would be at $\sim$ 60000 R$_\odot$ at the time of this first spectrum. From the lack of narrow lines, we can assume that any CSM outside this region was of sufficiently low density that, once ionised, it would recombine at a low enough rate to be undetectable. A steady stellar wind moving at $\sim$50 \kms~ would cover this space in about 30 years. So any stellar wind creating CSM around SN~2016gsd would need to be short lived or weak enough such that the CSM outside this region would be of low density. The dwarf host of SN~2016gsd points to a low metallicity progenitor, which is consistent with the weak metal lines in its spectra, and this could inhibit the existence of a strong pre$-$explosion stellar wind (although pre-SN mass loss might not be very dependent on metallicity, see e.g. \citet{Fuller2017}). It seems more likely therefore that the dense CSM around SN~2016gsd was created through a relatively short mass loss episode. This is consistent with the model of \citet{Dessart2016}, and has been suggested to occur around normal RSG progenitors \citep[see e.g.,][]{Yaron2017,Forster2018}.

In the +320 day spectrum of SN~2016gsd, we see a broad line profile with evidence for Lorentzian wings in the H$\alpha$ emission, which could indicate electron scattering in a dense CSM. Here we consider some scenarios in which such a line profile could be formed. Assuming a wind composed of pure H with density $\propto 1/r^2$, the number density of electrons $N_{e}$ is:
\begin{multline}
N_{e} = 3\times10^{8} \left(\frac{\dot{M}}{10^{-4}~\text{M}_{\odot}~\text{yr}^{-1}}\right) 
\\
\times \left(\frac{v_{\text{wind}}}{100~\text{km~s}^{-1}}\right) \left(\frac{R}{10^{15}~\text{cm}}\right)^{-2}\text{cm}^{-3}.   
\end{multline}
If we assume a line forms at the inner edge, $R_{s}$, of such a wind, the optical depth is: 
\begin{equation}
\label{equation:int_tau}
\tau_{\rm e} = \int_{R_{s}}^{\infty} \sigma_{\rm T} N_{e} dR
\end{equation}
where $\sigma_{\rm T}$ is the electron Thomson cross section. Assuming that the shock radius is $\sim 20\%$ further out than the fastest ejecta material, one can express $R_s$ in terms of the maximum ejecta velocity $v_{ej}$ and elapsed time. For effective electron scattering, we require $\tau_{\rm e} \geq 1$. Evaluating the integral in equation \ref{equation:int_tau} we find:
\begin{equation}
\tau_{\rm e} = 1 < \sigma_{\rm T} N_{e} R_{s}
\end{equation}
and substituting for $R_{s}$ leads to:
\begin{multline}
\dot{M} \geq 0.104 \left(\frac{v_{\text{wind}}}{100~\text{km~s}^{-1}}\right)
\\
\times \left(\frac{v_{\text{ej}}}{2\textsf{x}10^{4}~\text{km~s}^{-1}}\right)\left(\frac{t}{100~\text{d}}\right) \text{M}_{\odot}~\text{yr}^{-1}.
\end{multline}
Taking typical velocity values such as $v_{\text{wind}} =$~50 \kms~and $v_{\text{ej}} =$ 5\,000 \kms, our spectrum at t~$=$~300 rest frame days requires $\dot{M} \approx 0.038~M_{\odot}$ yr$^{-1}$ to have the required optical depth for scattering. This number seems too large to be feasible as in such a case the ejecta would be concealed behind an electron scattering shell even at early times, which we do not see, and the resulting SN would be much more luminous than SN~2016gsd.

An alternative possibility would be to achieve these densities only in a smaller region due to the geometric distribution in the CSM, i.e., a CSM disk, or clump. We assume the same as the above, with mass loss all taking place into a disk. Taking some disk of height $h$, the fraction of solid angle, $f$, it would cover would be:
\begin{equation}
f = \frac{2\pi R_{s}h}{4\pi R_{s}^{2}} = \frac{h}{2R_{s}} = \frac{x}{2} \quad \text{where} \quad x := \frac{h}{R_{s}}.
\end{equation}
If all mass loss is concentrated to a disk, the number density of electrons in the disk  is:
\begin{equation}
N_{e,\text{disk}} = \frac{N_{e,\text{wind}}}{f}.
\end{equation} 
The smallest optical depth should be perpendicular to the disk, and we can use $h$ as a typical path length. For optical depth greater than unity:
\begin{equation}
\tau_{\rm e} = 1 \leq \sigma_{\rm T} \left(\frac{2N_{e,\text{wind}}}{x}\right) x R_{s} = 2\sigma_{\rm T} N_{e,\text{wind}}R_{s}  
\end{equation}
which is the same as for the shell, apart from a factor of 2. Here we have assumed that fastest ejecta in the disk plane reach out to the disk, i.e., out to $R_{disk}$. Setting the optical depths in the two scenarios equal, we find:
\begin{equation}
\left(\frac{\dot{M}}{v}\right)_{\text{disk}} = \left(\frac{\dot{M}}{v}\right)_{\text{wind}}\frac{b}{2.4} \quad \text{where} \quad b:= \frac{R_{\text{disk}}}{R_{s,\text{wind}}}.
\end{equation}

The velocity of the fastest ejecta in the disk plane will be slower than in other directions where it can expand more freely, so $R_{\text{disk}} < R_{s,\text{wind}}$. Depending on the density distribution of the disk compared to the surrounding circumstellar gas, the inner radius of the shock in the disk can be much smaller than that in the CSM outside it \citep[e.g.,][]{MacD18}. As long as the disk mass is significantly less than the ejecta mass, then $b \sim f^{1/(n-2)}$ if the total mass loss in the disk and in the wind are equal. Here $n$ is the density slope of the unshocked supernova ejecta, i.e., $\rho_{ej} \propto R^{-n}$. Typically, $n \sim 10$, so with $f=0.01$, $b \approx 0.56$. A situation with $b<0.5$ would require a lower overall mass loss in the wind than in the disk. For $b=0.5$, 
we find:
\begin{equation}
\dot{M}_{\text{disk}} > 0.5\dot{M}_{\text{wind}} / 2.4 = 0.008~M_{\odot} ~\text{year}^{-1},
\end{equation}
to obtain $\tau_{\rm e} > 1$ for the typical parameters we used for the
wind case. In reality, a smaller value for $\dot{M}_{\text{disk}}$ could be acceptable as we have assumed no further electron scattering from the SN ejecta and any other CSM in our line of sight. The value we derive would still imply a very large mass loss rate. A point working against the existence of such a disk is that it would produce narrow H$\alpha$ emission at early times through recombination of H in the unshocked region of the disk, which is necessarily dense enough to produce such a feature. As SN 2016gsd appears rather like a ``normal" type II SN during the photospheric phase, which is inconsistent with the high mass loss rates we derive here, it seems unlikely that the line we observe is produced purely by electron scattering in dense CSM. For a more detailed analysis of CSM interaction with a disk, see e.g. \citet{Vlasis2016}.

Although this analysis disfavours electron scattering in dense CSM, there is other evidence of CSM interaction. As discussed in Sect. \ref{sect:Spectroscopy}, from 36$-$132d we see an intermediate width component in the H$\alpha$ P-Cygni profile of SN~2016gsd centred at the rest wavelength of H$\alpha$ whose FWHM evolves from $\sim$2000\kms~ to $\sim$3000\kms~ across the 100d period. This component is similar to those seen in the so-called 1988Z-like group of type IIn SNe such as SN 1988Z, SN 1995G and SN 2005ip \citep[see e.g.][]{Stathakis1991,Pastorello2002,Stritzinger2012}, where they are associated with radiation from post-shock gas. The evolution of the velocity derived from the FWHM of this component is not dissimilar to that found in SN 2005ip, as can be seen in Fig. 15 of \citet{Stritzinger2012}, implying a similar average velocity evolution of the post-shock gas. The luminosity of the component also increases during this period.

Our photometry at +132 and +179d implies a mass of synthesised $^{56}$Ni considerably larger than implied by our late-time spectroscopy. We can infer from this that the luminosity we see at +132 and 179d is not entirely powered by $^{56}$Ni and must be boosted by CSM interaction. Furthermore, the r band magnitude at 305d implies a large decrease in the SN decline rate ,although there is some uncertainty about this measurement. This is inconsistent with $^{56}$Ni power, that can only decline more quickly over time. As we have already discussed above, the model of \citet{Dessart2016} well explains the hot and mostly featureless spectrum that we see at +23 and +25d and requires the presence of a large amount of CSM. Thus we can see the influence of CSM throughout the evolution of the SN.

However, the high velocities of the SN ejecta and the presence of a strong H$\alpha$ absorption differentiate SN~2016gsd from SN~1998S and present some challenges for the model of \citet{Dessart2016}.  We see the H$\alpha$ absorption strongly throughout the +36$-$132d period at a velocity as high as $\sim$20000 \kms. This is exceptionally high for type II SNe (indicating a large explosion energy) and this material can not have been decelerated by interaction with dense CSM. Furthermore, the associated H$\alpha$ absorption is not filled in by emission from the CDS, but is instead clearly visible. In \citet{Moriya2011,Moriya2017}, the authors model the type II SN with dense CSM within 10$^{15}$cm and find that the photospheric velocity at early times is greatly lowered, due to the photosphere being within the unshocked dense CSM. Whether the photospheric velocity is affected after $\sim$25 days depends on the CSM mass. If there is more than $\sim$1M$_{\odot}$ of CSM, the suppression of velocities lasts longer, and the light curve is also affected at late times. For these models to achieve the peak luminosity of SN~2016gsd would require a large CSM mass which would suppress the velocities and be inconsistent with the high velocities we see in the spectrum of SN~2016gsd at +36d. 

These facts point towards asymmetry in the configuration of CSM. This would provide a line of sight to the SN ejecta that would allow us to see the broad SN features and would also allow the fast material to freely expand along this direction unhindered by interaction with the CSM. Interaction would still occur in other directions where there is dense CSM, providing the radiated energy that we observe. For example, a scenario where the interaction was happening with a dense CSM disk overrun and enveloped by the SN ejecta was proposed to explain the properties of the peculiar type II SN iPTF14hls and the transitional type II/IIn object iPTF11iqb \citet{Smith2015,Andrews18}. We do not see see asymmetry or wavelength shifts in the spectral features of SN 2016gsd associated with the CSM, namely the intermediate width H$\alpha$ component and the H$\alpha$ feature in the late time spectrum, which would also have indicated asymmetric CSM. Although we do not suggest any particular configuration for this CSM material here, the lack of these features in our data could provide constraints on the CSM configuration, especially given that such features are not uncommon in interacting SNe (see e.g. SN~1998S \citep{Fassia2001} and SN~2014G \citep{Terreran2016}).

\subsection{Possible progenitors for SN~2016gsd}
\label{sect:disc3}

There are some indications that type IIL SNe could result from higher mass progenitors than type IIP SNe. Evidence for interaction in bright type IIL SNe such as SNe 1979C and 1998S implies that they experienced larger mass loss. Furthermore, the requirement of an inflated and low mass H-rich envelope that is suggested to explain the brightness and linear light curve of such objects both require significant mass loss during the progenitor's life time. Models show that greater mass loss through stellar winds results from a larger ZAMS mass for the progenitor \citep{Heger2003,Kasen2009}. Constraints for the progenitor masses of SNe 1979C and 2013fc have been derived as ($17$-$18)\pm$3 M$_{\odot}$ \citep{VanDyk1999} and 19$\pm$4 M$_{\odot}$ \citep{Kangas2016} respectively, based on modelling of the stellar populations in the local environment of the SN. \cite{Fassia2001} suggest a massive (although <25 M$_{\odot}$) progenitor for SN 1998S based on the high rate of mass loss they infer from the CSM properties and the large He core mass implied by their modelling of their observed CO spectrum. Thus there is some evidence that these bright type IIL SNe come from a more massive population of RSGs than normal type IIs. 

In the case of SN~2016gsd, we can constrain the progenitor mass from the non$-$detection of [O~{\sc i}] $\lambda$6300,6364 in our +321d spectrum. The strength of this line in nebular spectra of Fe core-collapse SNe is strongly connected to the synthesised oxygen mass, which is itself strongly dependent on the ZAMS mass \citep{Woosley1995,Thielemann1996}. Fig. \ref{fig:late_spec} shows that we would expect to see this line in our spectrum, for a 15 M$_{\odot}$ progenitor, assuming 0.062 M$_{\odot}$ of synthesised $^{56}$Ni. As discussed in \cite{Jerkstrand2012,Jerkstrand2014}, the strength of this line is sensitive to increasing the ZAMS mass, and we would expect it to approximately double in strength as a percentage of the $^{56}$Co luminosity if we increased the ZAMS mass to 19 M$_{\odot}$. This argues against a progenitor initially more massive than 15 M$_{\odot}$.

We could also reduce the strength of the [O~{\sc i}] $\lambda$6300,6364 feature by reducing the mass of synthesised $^{56}$Ni. It would not be unusual for the mass of synthesised $^{56}$Ni produced by a type II SN to be much lower than 0.062 M$_{\odot}$. \citet{Muller2017} examined a sample of 38 type II SNe and found a median $^{56}$Ni mass of 0.031 M$_{\odot}$ and mean of 0.046 M$_{\odot}$. Similar to this, \citet{Anderson2019} found the median and mean $^{56}$Ni mass for 115 type II SNe from the literature to be 0.032 M$_{\odot}$ and 0.044 M$_{\odot}$ respectively. However, CCSN models yield a correlation between explosion energy and the mass of synthesised $^{56}$Ni \citep[see e.g.][]{Sukhbold2016} which is supported by the observations in the sample of \citet{Muller2017}. In particular, \citet{Sukhbold2016} find that stars more massive than 12 M$_{\odot}$ synthesise considerably more $^{56}$Ni than those below. The high velocities we observe in SN~2016gsd point towards a more energetic explosion, which precludes a significantly lower mass of synthesised $^{56}$Ni. An alternative method to reduce the mass of synthesised $^{56}$Ni that we can not rule out would be significant fallback onto the compact remnant after core-collapse. Such a model would require a black hole accretion wind to simultaneously produce the high velocities we observe \citep{Dexter2013}.

 The [O~{\sc i}] $\lambda$6300,6364 feature could be suppressed by collisional de$-$excitation if the densities in the core regions were high. However, there are many indications that the velocity of the ejecta in SN~2016gsd is higher than observed in other type II SNe: all measured velocities are much larger than is typical and modelling indicates that we require a partially stripped envelope, which should produce high velocities. Indeed, the velocity measured from the FWHM of the H$\alpha$ feature in the 321d spectrum is high, at $\sim$6000 \kms. Given the high velocities we observe, we would not expect core densities higher than typical for type II SNe.

\section{Conclusions}
\label{sect:Conclusions}
In this work we presented observations of the supernova~2016gsd. With a peak absolute magnitude of $V=$-$19.95$, this object is one of the brightest type II SNe and displays an exceptionally long and linear decline. The observed luminosity and line ratios of the host galaxy of the SN indicate a low metallicity progenitor, which is consistent with the weakness of the metal lines in its spectra. Comparison to a model produced with the new JEKYLL code indicates that the explosion of a 15 solar mass star with a depleted and inflated H-rich envelope can reproduce some of the characteristics of SN~2016gsd, but can not reproduce its high luminosity and extreme linearity. There are a number of observational signatures indicating significant influence from CSM interaction throughout the evolution of this object. At early times, the blue spectra with broad and weak emission features more than a week after the peak luminosity may point towards the emitting region being within a CDS which obscures the underlying SN features, as in the model of \citet{Dessart2016} for SN~1998S. Throughout the photospheric phase, the presence of an intermediate width H$\alpha$ emission feature implies ongoing emission from post shock gas. At late times, the presence of a continuum and prominent broad H$\alpha$ emission but the lack of forbidden nebular lines in our +320d spectrum imply continued interaction power. Finally, we derive a lower limit for the total radiated energy from the pseudo-bolometric light curve of 6.5x10$^{49}$ erg which implies the need for additional conversion of kinetic energy in the SN ejecta into radiation through interaction with a dense CSM. 

This SN joins the group of ``bright IILs'' with signs of interaction while showing some different characteristics. The inferred low metallicity progenitor of SN~2016gsd differentiates it from the other objects in this group, and thus may have played a role in determining its observed characteristics. It seems likely that the amount and geometry of the CSM around the progenitors of these SNe is the main determiner of their varied properties.  In the case of SN~2016gsd, the persistent very high velocities and strong absorption in H$\alpha$ point towards a significantly asymmetric configuration of CSM. Given how few SNe fall into this group, it will take more observations of such SNe to truly understand where they lie on the spectrum of progenitors from IIP$-$IIn supernovae. The ability of future deeper surveys to constrain the pre$-$SN outburst history of such objects will allow us to probe the allowed distributions of CSM that can be invoked in attempts to model them.

\section*{Acknowledgements}
We thank the anonymous referee for helpful comments and suggestions. We thank Massimo Turatto for observations taken at the Asiago Observatory, to N\'idia Morrell and Eric Hsiao for contributing to data reduction, to Anthony Piro for useful discussion and to Avishay Gal-Yam for helpful comments. T.R. acknowledges the financial support of the Jenny and Antti Wihuri foundation and the Vilho, Yrj{\"o} and Kalle V{\"a}is{\"a}l{\"a} Foundation of the Finnish academy of Science and Letters. We acknowledge the support of the UCD seed funding scheme (SF1518). M.F. acknowledges the support of a Royal Society $-$ Science Foundation Ireland University Research Fellowship. The JEKYLL simulations were performed on resources provided by the Swedish National Infrastructure for Computing (SNIC) at Parallelldatorcentrum (PDC). PL acknowledges support from the Swedish Research Council. M.S. is supported by a generous grant (13261) from VILLUM FONDEN and a  project grant (8021-00170B) from the Independent Research Fund Denmark (IRFD). NUTS2 is funded in part by the Instrument Center for Danish Astronomy (IDA). This work is based (in part) on observations collected at the European Organisation for Astronomical Research in the Southern Hemisphere, Chile as part of PESSTO (the Public ESO Spectroscopic Survey for Transient Objects) ESO program 188.D$-$3003, 191.D$-$0935, more ESO acknowledgements. The Pan$-$STARRS1 Surveys (PS1) and the PS1 public science archive have been made possible through contributions by the Institute for Astronomy, the University of Hawaii, the Pan$-$STARRS Project Office, the Max$-$Planck Society and its participating institutes, the Max Planck Institute for Astronomy, Heidelberg and the Max Planck Institute for Extraterrestrial Physics, Garching, The Johns Hopkins University, Durham University, the University of Edinburgh, the Queen's University Belfast, the Harvard$-$Smithsonian Center for Astrophysics, the Las Cumbres Observatory Global Telescope Network Incorporated, the National Central University of Taiwan, the Space Telescope Science Institute, the National Aeronautics and Space Administration under Grant No. NNX08AR22G issued through the Planetary Science Division of the NASA Science Mission Directorate, the National Science Foundation Grant No. AST$-$1238877, the University of Maryland, Eotvos Lorand University (ELTE), the Los Alamos National Laboratory, and the Gordon and Betty Moore Foundation. The SCUSS is funded by the Main Direction Program of Knowledge Innovation of Chinese Academy of Sciences (No. KJCX2$-$EW$-$T06). It is also an international cooperative project between National Astronomical Observatories, Chinese Academy of Sciences, and Steward Observatory, University of Arizona, USA. Technical support and observational assistance from the Bok telescope are provided by Steward Observatory. The project is managed by the National Astronomical Observatory of China and Shanghai Astronomical Observatory. Data resources are supported by Chinese Astronomical Data Center (CAsDC). S.D. and P.C. acknowledge Project 11573003 supported by NSFC. This research uses data obtained through the Telescope Access Program (TAP), which has been funded by the National Astronomical Observatories of China, the Chinese Academy of Sciences, and the Special Fund for Astronomy from the Ministry of Finance. SJS acknowledges STFC grant ST/P000312/1. This work has made use of data from the Asteroid Terrestial-impact Last Alert System (ATLAS) Prohect. ATLAS is primarily funded to search for near earth asteroids through NASA grants NN12AR55G, 80NSSC18K0284, and 80NSSC18K1575; byproducts of the NEO search include images and catalogs from the survey area. The ATLAS science products have been made possible through the contributions of the University of Hawaii Institute for Astronomy, the Queen's Univeristy Belfast, the Space Telescope Science Institute, and the South African Astronomical Observatory. O.R. acknowledge support by projects IC120009 ``Millennium Institute of Astrophysics (MAS)'' of the Iniciativa Cient\'ifica Milenio del Ministerio Econom\'ia, Fomento y Turismo de Chile and CONICYT PAI/INDUSTRIA 79090016. J.H. acknowledges financial support from the Finnish Cultural Foundation. Some data were taken with the Las Cumbres Observatory Network. G.H and D.A.H are supported by NSF grant AST-1313484. G.H thanks the LSSTC Data Science Fellowship Program, which is funded by LSSTC, NSF Cybertraining Grant \#1829740, the Brinson Foundation, and the Moore Foundation; his participation in the program has benefited this work. L.G. was funded by the European Union's Horizon 2020 research and innovation programme under the Marie Sk\l{}odowska-Curie grant agreement No. 839090. This work also makes use of observations collected at the European Southern Observatory under ESO programme 0103.D-0338(A). C.P.G. acknowledges support from EU/FP7-ERC grant no. [615929].



\bibliographystyle{mnras}
\bibliography{SN2016gsd_refs} 



\appendix
\section{Data Tables}

\begin{table*}
\centering
\resizebox{1\textwidth}{!}{%
\begin{tabular}{|c|c|c|c|c|c|c|c|c|c|c|c|c|} \hline
Star&RA (deg) &Dec (deg)&U&B&V&R&I&u&g&r&i&z\\ \hline
1 & 40.11470 & 19.20881 & 14.93$ \pm $0.17 & 14.77$ \pm $0.03 & 14.02$ \pm $0.01 & 13.57$ \pm $0.02 & 13.34$ \pm $0.02 & 15.69$ \pm $0.002 & 14.36$ \pm $0.01 & 13.78$ \pm $0.01 & 13.57$ \pm $0.01 & 13.48$ \pm $0.01 \\ 
2 & 40.18571 & 19.32015 & 15.06$ \pm $0.17 & 14.88$ \pm $0.03 & 14.16$ \pm $0.01 & 13.73$ \pm $0.02 & 13.50$ \pm $0.02 & 15.80$ \pm $0.003 & 14.49$ \pm $0.01 & 13.94$ \pm $0.01 & 13.78$ \pm $0.01 & 13.69$ \pm $0.01 \\ 
3 & 40.13688 & 19.32408 & 17.32$ \pm $0.16 & 17.35$ \pm $0.03 & 16.70$ \pm $0.01 & 16.32$ \pm $0.02 & 16.08$ \pm $0.02 & 18.10$ \pm $0.01 & 16.98$ \pm $0.01 & 16.51$ \pm $0.01 & 16.34$ \pm $0.01 & 16.29$ \pm $0.01 \\ 
4 & 40.10531 & 19.31996 & 15.56$ \pm $0.16 & 15.50$ \pm $0.03 & 14.84$ \pm $0.01 & 14.45$ \pm $0.02 & 14.22$ \pm $0.02 & 16.32$ \pm $0.003 & 15.13$ \pm $0.01 & 14.65$ \pm $0.01 & 14.51$ \pm $0.01 & 14.44$ \pm $0.01 \\ 
5 & 40.15990 & 19.23951 & 15.37$ \pm $0.17 & 15.26$ \pm $0.04 & 14.57$ \pm $0.02 & 14.17$ \pm $0.02 & 13.93$ \pm $0.02 & 16.12$ \pm $0.003 & 14.88$ \pm $0.01 & 14.36$ \pm $0.01 & 14.21$ \pm $0.02 & 14.12$ \pm $0.01 \\ 
6 & 40.12218 & 19.32932 & 17.88$ \pm $0.22 & 17.25$ \pm $0.04 & 16.29$ \pm $0.02 & 15.74$ \pm $0.02 & 15.49$ \pm $0.02 & 18.57$ \pm $0.01 & 16.76$ \pm $0.02 & 15.97$ \pm $0.01 & 15.72$ \pm $0.01 & 15.59$ \pm $0.01 \\ 
7 & 40.18775 & 19.32767 & $-$ & 19.76$ \pm $0.04 & 18.14$ \pm $0.02 & 17.24$ \pm $0.02 & 16.98$ \pm $0.02 & $-$ & 19.02$ \pm $0.02 & 17.55$ \pm $0.01 & 16.93$ \pm $0.01 & 16.56$ \pm $0.01 \\ 
8 & 40.19277 & 19.31007 & $-$ & 19.30$ \pm $0.04 & 17.95$ \pm $0.02 & 17.19$ \pm $0.02 & 16.94$ \pm $0.02 & $-$ & 18.67$ \pm $0.01 & 17.47$ \pm $0.01 & 17.07$ \pm $0.01 & 16.82$ \pm $0.01 \\ 
9 & 40.19461 & 19.26197 & $-$ & 17.99$ \pm $0.04 & 16.38$ \pm $0.01 & 15.47$ \pm $0.02 & 15.21$ \pm $0.02 & $-$ & 17.26$ \pm $0.01 & 15.78$ \pm $0.01 & 15.00$ \pm $0.01 & 14.56$ \pm $0.01 \\ 
10 & 40.12215 & 19.32561 & 16.72$ \pm $0.16 & 16.69$ \pm $0.03 & 16.02$ \pm $0.01 & 15.62$ \pm $0.02 & 15.38$ \pm $0.02 & 17.49$ \pm $0.01 & 16.31$ \pm $0.01 & 15.82$ \pm $0.01 & 15.66$ \pm $0.01 & 15.60$ \pm $0.01 \\ 
11 & 40.15227 & 19.25398 & $-$ & 19.40$ \pm $0.05 & 18.40$ \pm $0.04 & 17.83$ \pm $0.04 & 17.59$ \pm $0.04 & $-$ & 18.90$ \pm $0.02 & 18.07$ \pm $0.02 & 17.80$ \pm $0.01 & 17.63$ \pm $0.01 \\ 
12 & 40.09910 & 19.31565 & 19.00$ \pm $0.18 & 18.85$ \pm $0.04 & 18.12$ \pm $0.02 & 17.69$ \pm $0.03 & 17.45$ \pm $0.03 & 19.76$ \pm $0.03 & 18.45$ \pm $0.02 & 17.89$ \pm $0.01 & 17.70$ \pm $0.01 & 17.63$ \pm $0.01 \\ 
13 & 40.14182 & 19.32561 & $-$ & 20.18$ \pm $0.04 & 19.08$ \pm $0.03 & 18.45$ \pm $0.03 & 18.20$ \pm $0.03 & $-$ & 19.64$ \pm $0.02 & 18.70$ \pm $0.01 & 18.36$ \pm $0.01 & 18.18$ \pm $0.01 \\ 
14 & 40.12431 & 19.30970 & $-$ & 20.65$ \pm $0.05 & 20.07$ \pm $0.03 & 19.72$ \pm $0.04 & 19.48$ \pm $0.04 & $-$ & 20.31$ \pm $0.03 & 19.90$ \pm $0.02 & 19.72$ \pm $0.02 & 19.71$ \pm $0.05 \\ 
15 & 40.15600 & 19.30788 & $-$ & 21.12$ \pm $0.07 & 20.49$ \pm $0.06 & 20.12$ \pm $0.06 & 19.89$ \pm $0.06 & $-$ & 20.76$ \pm $0.05 & 20.31$ \pm $0.03 & 20.16$ \pm $0.02 & 20.17$ \pm $0.06 \\ 
16 & 40.15908 & 19.30376 & $-$ & 21.66$ \pm $0.06 & 20.13$ \pm $0.03 & 19.26$ \pm $0.04 & 19.01$ \pm $0.04 & $-$ & 20.96$ \pm $0.04 & 19.57$ \pm $0.01 & 19.10$ \pm $0.01 & 18.79$ \pm $0.02 \\ 
17 & 40.14221 & 19.27672 & $-$ & 21.21$ \pm $0.07 & 20.54$ \pm $0.07 & 20.14$ \pm $0.07 & 19.90$ \pm $0.07 & $-$ & 20.83$ \pm $0.05 & 20.34$ \pm $0.04 & 20.13$ \pm $0.03 & 19.99$ \pm $0.08 \\ 
18 & 40.11065 & 19.27469 & $-$ & 21.25$ \pm $0.05 & 19.55$ \pm $0.03 & 18.59$ \pm $0.03 & 18.34$ \pm $0.03 & $-$ & 20.48$ \pm $0.03 & 18.92$ \pm $0.01 & 18.00$ \pm $0.01 & 17.51$ \pm $0.01 \\ 
19 & 40.14780 & 19.25656 & $-$ & 19.95$ \pm $0.05 & 19.15$ \pm $0.03 & 18.68$ \pm $0.03 & 18.44$ \pm $0.03 & $-$ & 19.52$ \pm $0.03 & 18.89$ \pm $0.01 & 18.63$ \pm $0.01 & 18.47$ \pm $0.02 \\ 
20 & 40.13250 & 19.27764 & $-$ & 22.09$ \pm $0.07 & 20.41$ \pm $0.06 & 19.47$ \pm $0.06 & 19.22$ \pm $0.06 & $-$ & 21.33$ \pm $0.05 & 19.80$ \pm $0.03 & 19.04$ \pm $0.02 & 18.57$ \pm $0.02 \\ 
\end{tabular}%
}
\caption{Magnitudes for the calibration stars shown in Figure \ref{fig:finder}. Due to the limited standard star observations available, the \textit{BVRI} magnitudes were obtained by converting from Pan$-$STARRS1 magnitudes while U band magnitudes were obtained from the SCUSS survey. \textit{UBVRI} band magnitudes are in the Vega system and \textit{ugriz} band magnitudes are in the AB system.
\label{tab:calibration}}
\end{table*}

\begin{table*}
\centering
\begin{tabular}{|c|c|c|c|c|c|c|c|c|} \hline
Phase (days)&Date (UT)&MJD&Telescope&U&B&V&R&I\\ \hline
15.3 & 2016-10-02 & 57663.3 & LCO & - & 17.93 $ \pm $ 0.04 & 17.83 $ \pm $ 0.02 & - & - \\  
16.9 & 2016-10-03 & 57664.9 & LCO & - & 17.93 $ \pm $ 0.04 & 17.77 $ \pm $ 0.05 & - & - \\  
19.2 & 2016-10-06 & 57667.2 & LCO & - & 18.05 $ \pm $ 0.03 & 17.85 $ \pm $ 0.03 & - & - \\  
21.1 & 2016-10-08 & 57669.1 & LCO & - & 17.98 $ \pm $ 0.04 & 17.83 $ \pm $ 0.05 & - & - \\  
22.6 & 2016-10-09 & 57670.6 & LCO & - & - & 17.92 $ \pm $ 0.04 & - & - \\ 
24.0 & 2016-10-11 & 57672.0 & LCO & 17.51 $ \pm $ 0.03 & 18.12 $ \pm $ 0.02 & 17.92 $ \pm $ 0.01 & - & - \\  
25.1 & 2016-10-12 & 57673.1 & NOT & - & - & 17.94 $ \pm $ 0.01 & - & - \\  
25.2 & 2016-10-12 & 57673.2 & LCO & - & 18.19 $ \pm $ 0.03 & 17.93 $ \pm $ 0.02 & - & - \\  
27.5 & 2016-10-14 & 57675.5 & LCO & 17.71 $ \pm $ 0.24 & 18.23 $ \pm $ 0.11 & 17.99 $ \pm $ 0.14 & - & - \\  
31.1 & 2016-10-18 & 57679.1 & NOT & - & 18.49 $ \pm $ 0.05 & 18.11 $ \pm $ 0.02 & - & - \\  
32.6 & 2016-10-19 & 57680.6 & LCO & 17.95 $ \pm $ 0.10 & 18.46 $ \pm $ 0.06 & 18.19 $ \pm $ 0.07 & - & - \\  
34.1 & 2016-10-21 & 57682.1 & NTT & - & - & 18.28 $ \pm $ 0.05 & - & - \\  
35.1 & 2016-10-22 & 57683.1 & LCO & 18.14 $ \pm $ 0.11 & 18.60 $ \pm $ 0.07 & 18.22 $ \pm $ 0.08 & - & - \\  
38.0 & 2016-10-25 & 57686.0 & LCO & 18.26 $ \pm $ 0.20 & 18.76 $ \pm $ 0.07 & 18.30 $ \pm $ 0.07 & - & - \\  
41.2 & 2016-10-28 & 57689.2 & LCO & 18.54 $ \pm $ 0.05 & 18.92 $ \pm $ 0.06 & 18.43 $ \pm $ 0.04 & - & - \\  
43.9 & 2016-10-30 & 57691.9 & LCO & 18.61 $ \pm $ 0.06 & 19.08 $ \pm $ 0.05 & 18.53 $ \pm $ 0.03 & - & - \\  
46.5 & 2016-11-02 & 57694.5 & LCO & - & 19.19 $ \pm $ 0.04 & 18.69 $ \pm $ 0.06 & - & - \\  
47.9 & 2016-11-03 & 57695.9 & LCO & - & 19.35 $ \pm $ 0.04 & 18.77 $ \pm $ 0.02 & - & - \\  
49.0 & 2016-11-04 & 57697.0 & LCO & - & 19.43 $ \pm $ 0.04 & 18.70 $ \pm $ 0.07 & - & - \\  
50.0 & 2016-11-05 & 57698.0 & LCO & - & 19.53 $ \pm $ 0.11 & 18.79 $ \pm $ 0.06 & - & - \\  
53.6 & 2016-11-09 & 57701.6 & LCO & - & 19.61 $ \pm $ 0.05 & 18.83 $ \pm $ 0.06 & - & - \\  
59.8 & 2016-11-15 & 57707.8 & LCO & - & - & 19.11 $ \pm $ 0.25 & - & - \\  
62.8 & 2016-11-18 & 57710.8 & LCO & - & 20.10 $ \pm $ 0.06 & 19.07 $ \pm $ 0.05 & - & - \\  
63.8 & 2016-11-19 & 57711.8 & LCO & - & 20.21 $ \pm $ 0.06 & 19.06 $ \pm $ 0.06 & - & - \\  
68.0 & 2016-11-23 & 57716.0 & NOT & - & 20.40 $ \pm $ 0.05 & 19.23 $ \pm $ 0.03 & - & - \\  
68.8 & 2016-11-24 & 57716.8 & LCO & - & - & 19.44 $ \pm $ 0.11 & - & - \\  
69.2 & 2016-11-25 & 57717.2 & NTT & - & - & 19.35 $ \pm $ 0.10 & - & - \\  
72.2 & 2016-11-28 & 57720.2 & LCO & - & 20.68 $ \pm $ 0.06 & 19.47 $ \pm $ 0.06 & - & - \\  
74.8 & 2016-11-30 & 57722.8 & LCO & - & 20.86 $ \pm $ 0.07 & 19.56 $ \pm $ 0.05 & - & - \\  
78.1 & 2016-12-04 & 57726.1 & LCO & - & 20.88 $ \pm $ 0.06 & 19.60 $ \pm $ 0.06 & - & - \\  
79.9 & 2016-12-05 & 57727.9 & Asiago & - & 21.43 $ \pm $ 0.15 & 19.67 $ \pm $ 0.08 & - & - \\  
81.1 & 2016-12-07 & 57729.1 & LCO & - & 21.10 $ \pm $ 0.22 & 19.62 $ \pm $ 0.12 & - & - \\  
93.1 & 2016-12-19 & 57741.1 & LCO & - & 21.81 $ \pm $ 0.16 & 20.12 $ \pm $ 0.08 & - & - \\  
102.1 & 2016-12-28 & 57750.1 & NTT & - & 22.28 $ \pm $ 0.08 & 20.40 $ \pm $ 0.12 & 19.46 $ \pm $ 0.11 & - \\  
105.8 & 2016-12-31 & 57753.8 & LCO & - & - & 20.60 $ \pm $ 0.11 & - & - \\  
108.4 & 2017-01-03 & 57756.4 & LCO & - & - & 20.64 $ \pm $ 0.24 & - & - \\  
131.9 & 2017-01-26 & 57779.9 & NOT & - & 23.13 $ \pm $ 0.10 & 21.74 $ \pm $ 0.06 & - & - \\  
178.9 & 2017-03-14 & 57826.9 & NOT & - & - & 22.74 $ \pm $ 0.24 & - & 20.92 $ \pm $ 0.22 \\  
\hline
\end{tabular}
\caption{Photometry of SN~2016gsd in the \textit{UBVRI} bands. Magnitudes are given in the Vega system. Phases are relative to the explosion epoch obtained in Sect. \ref{subsec:Explosion epoch}.}
\label{tab:UBVRI_phot}
\end{table*}

\begin{table*}
\centering
\begin{tabular}{|c|c|c|c|c|c|c|c|c|} \hline
Phase (days)&Date (UT)&MJD&Telescope&u&g&r&i&z\\ \hline
15.3 & 2016-10-02 & 57663.3 & LCO & - & - & 17.84 $ \pm $ 0.02 & 17.95 $ \pm $ 0.03 & - \\  
16.9 & 2016-10-03 & 57664.9 & LCO & - & - & 17.80 $ \pm $ 0.02 & 17.96 $ \pm $ 0.05 & - \\  
19.2 & 2016-10-06 & 57667.2 & LCO & - & - & 17.94 $ \pm $ 0.03 & 17.95 $ \pm $ 0.03 & - \\  
21.1 & 2016-10-08 & 57669.1 & LCO & - & - & 17.85 $ \pm $ 0.02 & 18.01 $ \pm $ 0.04 & - \\  
22.6 & 2016-10-09 & 57670.6 & LCO & - & - & 17.93 $ \pm $ 0.03 & 17.91 $ \pm $ 0.05 & - \\ 
24.0 & 2016-10-11 & 57672.0 & LCO & - & 17.99 $ \pm $ 0.02 & 17.92 $ \pm $ 0.03 & 17.91 $ \pm $ 0.03 & - \\  
25.1 & 2016-10-12 & 57673.1 & NOT & - & 17.95 $ \pm $ 0.01 & - & 17.95 $ \pm $ 0.01 & 18.10 $ \pm $ 0.01 \\  
25.2 & 2016-10-12 & 57673.2 & LCO & - & - & 17.92 $ \pm $ 0.03 & 17.97 $ \pm $ 0.03 & - \\  
27.6 & 2016-10-14 & 57675.6 & LCO & - & 18.12 $ \pm $ 0.10 & 17.98 $ \pm $ 0.06 & 17.96 $ \pm $ 0.04 & - \\  
28.7 & 2016-10-15 & 57676.7 & LCO & - & - & 17.89 $ \pm $ 0.12 & 18.02 $ \pm $ 0.10 & - \\  
31.1 & 2016-10-18 & 57679.1 & NOT & 18.65 $ \pm $ 0.06 & 18.19 $ \pm $ 0.01 & 18.07 $ \pm $ 0.03 & 17.96 $ \pm $ 0.03 & 17.94 $ \pm $ 0.04 \\  
32.6 & 2016-10-19 & 57680.6 & LCO & - & - & 18.12 $ \pm $ 0.04 & 18.10 $ \pm $ 0.03 & - \\  
35.7 & 2016-10-22 & 57683.7 & LCO & - & 18.50 $ \pm $ 0.03 & 18.10 $ \pm $ 0.06 & 18.09 $ \pm $ 0.08 & - \\  
38.9 & 2016-10-25 & 57686.9 & LCO & - & 18.57 $ \pm $ 0.02 & 18.26 $ \pm $ 0.06 & 18.26 $ \pm $ 0.03 & - \\  
41.2 & 2016-10-28 & 57689.2 & LCO & - & 18.79 $ \pm $ 0.02 & 18.30 $ \pm $ 0.04 & 18.25 $ \pm $ 0.04 & - \\  
43.9 & 2016-10-30 & 57691.9 & LCO & - & 18.87 $ \pm $ 0.02 & 18.40 $ \pm $ 0.03 & 18.26 $ \pm $ 0.03 & - \\  
46.5 & 2016-11-02 & 57694.5 & LCO & - & - & - & 18.36 $ \pm $ 0.02 & - \\  
48.0 & 2016-11-03 & 57696.0 & LCO & - & - & 18.56 $ \pm $ 0.02 & 18.34 $ \pm $ 0.02 & - \\  
49.0 & 2016-11-05 & 57697.0 & LCO & - & 18.99 $ \pm $ 0.05 & 18.53 $ \pm $ 0.03 & 18.25 $ \pm $ 0.03 & - \\  
53.6 & 2016-11-09 & 57701.6 & LCO & - & 19.34 $ \pm $ 0.05 & 18.68 $ \pm $ 0.03 & 18.40 $ \pm $ 0.02 & - \\  
59.8 & 2016-11-15 & 57707.8 & LCO & - & - & 18.72 $ \pm $ 0.14 & 18.51 $ \pm $ 0.07 & - \\  
62.8 & 2016-11-18 & 57710.8 & LCO & - & 19.72 $ \pm $ 0.03 & 18.86 $ \pm $ 0.02 & 18.59 $ \pm $ 0.03 & - \\  
63.8 & 2016-11-19 & 57711.8 & LCO & - & - & 18.90 $ \pm $ 0.02 & 18.55 $ \pm $ 0.05 & - \\  
68.0 & 2016-11-23 & 57716.0 & NOT & 21.79 $ \pm $ 0.13 & 20.05 $ \pm $ 0.02 & 18.97 $ \pm $ 0.03 & 18.59 $ \pm $ 0.03 & 18.70 $ \pm $ 0.03 \\  
68.8 & 2016-11-24 & 57716.8 & LCO & - & 20.01 $ \pm $ 0.04 & 19.02 $ \pm $ 0.04 & 18.70 $ \pm $ 0.06 & - \\  
72.2 & 2016-11-28 & 57720.2 & LCO & - & - & 19.15 $ \pm $ 0.03 & 18.75 $ \pm $ 0.04 & - \\  
74.9 & 2016-11-30 & 57722.9 & LCO & - & 20.23 $ \pm $ 0.06 & 19.19 $ \pm $ 0.01 & 18.73 $ \pm $ 0.03 & - \\  
78.1 & 2016-12-04 & 57726.1 & LCO & - & - & 19.23 $ \pm $ 0.03 & 18.93 $ \pm $ 0.02 & - \\  
79.9 & 2016-12-05 & 57727.9 & Asiago & - & 20.57 $ \pm $ 0.09 & 19.26 $ \pm $ 0.05 & 18.98 $ \pm $ 0.02 & 18.93 $ \pm $ 0.09 \\  
81.1 & 2016-12-07 & 57729.1 & LCO & - & 20.49 $ \pm $ 0.10 & 19.32 $ \pm $ 0.03 & 18.89 $ \pm $ 0.04 & - \\  
93.1 & 2016-12-19 & 57741.1 & LCO & - & 21.02 $ \pm $ 0.06 & 19.69 $ \pm $ 0.03 & 19.11 $ \pm $ 0.04 & - \\  
102.2 & 2016-12-28 & 57750.2 & NTT & - & - & - & 19.44 $ \pm $ 0.12 & - \\  
105.8 & 2016-12-31 & 57753.8 & LCO & - & 21.62 $ \pm $ 0.15 & 19.98 $ \pm $ 0.11 & 19.29 $ \pm $ 0.05 & - \\  
108.5 & 2017-01-03 & 57756.5 & LCO & - & - & 19.85 $ \pm $ 0.12 & - & - \\  
109.8 & 2017-01-04 & 57757.8 & LCO & - & - & 19.96 $ \pm $ 0.08 & 19.54 $ \pm $ 0.05 & - \\  
117.3 & 2017-01-12 & 57765.3 & LCO & - & - & - & 19.56 $ \pm $ 0.14 & - \\  
131.2 & 2017-01-26 & 57779.2 & LCO & - & - & 20.87 $ \pm $ 0.15 & - & - \\  
131.9 & 2017-01-26 & 57779.9 & NOT & - & - & 20.98 $ \pm $ 0.04 & 20.25 $ \pm $ 0.02 & - \\  
178.9 & 2017-03-14 & 57826.9 & NOT & - & - & 21.85 $ \pm $ 0.08 & - & - \\  
305.2 & 2017-07-19 & 57953.2 & NOT & - & - & 22.77 $ \pm $ 0.24 & - & - \\  
\hline
\end{tabular}
\caption{Photometry of SN~2016gsd in the \textit{ugriz} bands. Magnitudes are given in the AB system. Phases are relative to the explosion epoch obtained in Sect. \ref{subsec:Explosion epoch}.}
\label{tab:ugriz_phot}
\end{table*}

\begin{table*}
\centering
\begin{tabular}{|c|c|c|c|c|c|c|c|c|} \hline
Phase (days)&Date (UT)&MJD&Instrument&J&H&K\\ \hline
57.1 & 2016$-$11$-$13 & 57705.1 &  NOTCam  & 17.88$ \pm $ 0.06 & 17.62$ \pm $ 0.09 & 17.42$ \pm $ 0.09 \\ 
68.6 & 2016$-$11$-$24 & 57716.6 &  SOFI  & 17.96$ \pm $ 0.06& 17.73$ \pm $ 0.09 & $-$ \\ 
89.9 & 2016$-$12$-$15 & 57737.9 &  NOTCam  & 18.13$ \pm $ 0.07 & 18.03$ \pm $ 0.06 & 17.49$ \pm $ 0.08 \\ 
115.0 & 2017$-$01$-$09 & 57763.0 &  NOTCam  & 18.89$ \pm $ 0.05 & 18.31$ \pm $ 0.08 & 18.06$ \pm $ 0.08 \\ 
352.2 & 2017$-$09$-$04 & 58000.2 &  NOTCam  & $-$ & $-$ &  > 19.83 \\ 
\hline
\end{tabular}
\caption{Photometry of SN~2016gsd in the JHK bands. Magnitudes are given in the Vega system. Phases are relative to the explosion epoch obtained in Sect. \ref{subsec:Explosion epoch}}.
\label{tab:JHK_phot}
\end{table*}

\begin{table*}
\centering
\begin{tabular}{|c|c|c|c|c|c|c|} \hline
Phase (observer days) & Date (UT) & MJD& Telescope & Instrument & Grism & Range (\r{A})\\ \hline
 23.1 & 2016-10-10 & 57671.1  & NOT & ALFOSC & Gr4  & 3200$-$9600  \\      
 25.1 & 2016-10-12 & 57673.1  & NOT & ALFOSC & Gr4  & 3200$-$9600  \\
25.7 & 2016-10-12 & 57673.7  & Baade & FIRE & $-$  & 8200$-$25100  \\  
 34.1 & 2016-10-21 & 57682.1  & NTT & EFOSC2 & Gr13  & 3650$-$9246  \\      
 44.1 & 2016-10-31 & 57692.1  & NOT & ALFOSC & Gr4  & 3200$-$9600  \\
 46.2 & 2016-11-02 & 57694.2  & NTT & EFOSC2 & Gr13  & 3650$-$9246  \\ 
 67.2  & 2016-11-22 & 57715.2  & NTT & EFOSC2 & Gr13  & 3650$-$9246  \\   
 69.2 & 2016-11-24 & 57717.2  & NTT & EFOSC2 & Gr13  & 3650$-$9246  \\  
 102.1 & 2016-12-28 & 57750.1  & NTT & EFOSC2 & Gr13  & 3650$-$9246  \\ 
 131.9 & 2017-01-26 & 57779.9  & NOT & ALFOSC & Gr4  & 3200$-$9600  \\ 
 321.1 & 2017-08-04 & 57969.1  & GTC & OSIRIS & R1000B  & 3630$-$7850 \\ 

\end{tabular}
\caption{Log of spectroscopic observations. Phases are relative to the explosion epoch obtained in Sect. \ref{subsec:Explosion epoch}}.
\label{tab:Spectral log}
\end{table*}

  


\bsp	
\label{lastpage}
\end{document}